\documentclass[a4paper,fleqn,usenatbib]{mnras}

\usepackage{newtxtext,newtxmath}

\usepackage[T1]{fontenc}
\usepackage{ae,aecompl}

\usepackage{graphicx}	
\usepackage{amsmath}	
\usepackage{amssymb}	

\usepackage{threeparttable}

\newcommand{\chis}{$\chi^{2}$}	
\newcommand{\swift}{\textit{Swift}}
\newcommand{\xmm}{\textit{XMM-Newton}}
\newcommand{\cnts}{\mathrm{counts~s}^{-1}}
\newcommand{\flux}{\mathrm{erg~cm}^{-2}~\mathrm{s}^{-1}}

\title[The X-ray transient MAXI~J1807+132]{The complex evolution of the X-ray binary transient MAXI~J1807+132 along the decay of its discovery outburst}

\author[F. Jim\'{e}nez-Ibarra et al.]{F. Jim\'{e}nez-Ibarra,$^{1,2}$\thanks{E-mail: felipeji@iac.es}
T. Mu\~{n}oz-Darias,$^{1,2}$
M. Armas Padilla,$^{1,2}$
D. M. Russell,$^{3}$\newauthor
J. Casares,$^{1,2}$
M.A.P. Torres,$^{1,2}$
D. Mata S\'{a}nchez,$^{1,2,4,5}$
P. G. Jonker$^{4,6}$ and
F. Lewis$^{7,8}$ 
\\
$^{1}$Instituto de Astrof\'{i}sica de Canarias, V\'{i}a L\'{a}ctea, La Laguna, E-38205, Santa Cruz de Tenerife, Spain\\
$^{2}$Departamento de Astrof\'{i}sica, Universidad de La Laguna, E-38206, Santa Cruz de Tenerife, Spain\\
$^{3}$New York University Abu Dhabi, PO Box 129188, Abu Dhabi, UAE \\
$^{4}$SRON, Netherlands Institute for Space Research, Sorbonnelaan 2, NL-3584CA Utrecht, The Netherlands \\
$^{5}$Jodrell Bank Centre for Astrophysics, School of Physics and Astronomy, The University of Manchester, Manchester M13 9PL, UK \\
$^{6}$Department of Astrophysics/IMAPP, Radboud University, P.O. Box 9010, NL-6500GL Nijmegen, The Netherlands \\
$^{7}$Faulkes Telescope Project, School of Physics, and Astronomy, Cardiff University, The Parade, Cardiff, CF24 3AA, Wales, UK\\
$^{8}$Astrophysics Research Institute, Liverpool John Moores University, 146 Brownlow Hill, Liverpool L3 5RF, UK
}

\date{Accepted XXX. Received YYY; in original form ZZZ}

\pubyear{2018}

\begin{document}
\label{firstpage}
\pagerange{\pageref{firstpage}--\pageref{lastpage}}
\maketitle

\begin{abstract}
MAXI~J1807+132 is an X-ray transient discovered during the decay of an outburst in 2017. We present optical and X-ray monitoring of the source over more than 125 days, from outburst to quiescence. The outburst decay is characterized by the presence of several re-flares with a quasi-periodic recurrence time of $\sim 6.5$ days. We detect broad H and He emission lines during outburst, characteristic of transient low mass X-ray binaries. These emission lines show strong variability from epoch to epoch and, in particular, during the early stages are found embedded into deep and very broad absorption features. The quiescent spectrum shows H$\alpha$ in emission and no obvious signatures of the donor star. \xmm\ and \swift\ spectra can be fitted with standard X-ray models for accreting black-holes and neutron stars, although the obtained spectral parameters favour the latter scenario.  Conversely, other observables such as the optical/X-ray flux ratio, the likely systemic velocity ($\gamma \sim -150$~km~s$^{-1}$) and the re-flares recurrence time suggest  a black hole nature. We discuss all the above possibilities with emphasis on the strong similarities of MAXI~J1807+132 with short orbital period systems.
\end{abstract}




\section{Introduction}

About a million X-ray sources have been detected after more than 50 years of X-ray astronomy. A fraction of these objects have been identified as low-mass X-ray binaries (LMXBs); binary systems harbouring a neutron star (NS) or a black hole (BH), which is accreting mass from a companion star typically less massive than the Sun. LMXBs provide a unique scenario to study extreme astrophysical phenomena such as accretion processes, the ejection of outflows and the final stages of the stellar evolution \citep[e.g.,][]{Fender2016,Casares2014}. The subclass known as transient LMXBs spend most part of their lives in a dormant, quiescent state, but show brightening episodes (outbursts) when they increase their luminosity by several orders of magnitude \citep[e.g.,][]{Remillard2006}. A significant fraction of the X-ray photons are absorbed in the accretion disc and then re-emitted at lower energies. As a result, the optical spectra of LMXBs are typically flat and blue with strong superimposed emission lines of H, \ion{He}{i}, and \ion{He}{ii} \citep[e.g.,][]{Charles2006}. 

X-ray and optical observations provide complementary information in the study of LMXBs. On one hand, the nature of the compact object and its properties can be constrained based on the presence of thermonuclear burst \citep{Galloway2008}, pulsations and other timing features \citep{vanderKlis2006,Motta2016}, and the spectral/timing X-ray evolution \citep{Belloni2011}. On the other hand, dynamical solutions can be obtained from optical and near-infrared spectroscopy during quiescence \citep[e.g.,][]{Casares1992}. In addition, the inflow and outflow properties (e.g., winds) as well as some scale parameters are accessible when high-quality spectroscopy is achievable in the optical \citep[e.g.,][]{Munoz-Darias2016,Jimenez-Ibarra2018} or in X-rays \citep[e.g.,][]{Miller2006, DiazTrigo2006, Ponti2012}.

MAXI~J1807+132 is a new X-ray transient discovered on 13 March 2017 by the nova-search system of the Monitor of All-sky X-ray Image \citep[MAXI,][]{Negoro2017}. The source is located 15 $^{\circ}$ above the galactic plane at $\alpha$, $\delta$ = $18^{ \mathrm{h}}08^{\mathrm{m}}07.549^{\mathrm{s}}$, $+13^{\circ}15^{\prime} 05.40^{\prime\prime} $ \citep[J2000,][]{Kennea2017a,Kennea2017b}. An optical counterpart consistent with the position of the transient was found in pre-outburst archival images of PanSTARRS-1 \citep{Chambers2016}.The object was identified in 31 multi-epoch images in 5 PanSTARRS-1 broadband filters: \textit{g}, \textit{r}, \textit{i}, \textit{z}, and \textit{y} \citep{Denisenko2017}. The average magnitude is \textit{g}~$\sim21$ but shows variability of up to 1 magnitude from epoch to epoch ( $ \sigma\sim0.8$ over 9 detections). Based on its high galactic latitude, soft X-ray spectrum, UV brightness, and a tentative association with a previous flaring event it was initially proposed as a candidate tidal disruption event \citep{Negoro2017,Kennea2017}. However, the optical spectrum and other X-ray properties soon advocated for an X-ray binary association \citep{Munoz-Darias2017a,Shidatsu2017a,ArmasPadilla2017a}. Subsequently, \citet{Shidatsu2017} favoured a NS accretor from analysis of \textit{Swift} X-ray spectra in the range of $0.3-10$ keV.  

In this paper we present a detailed optical and X-ray study of MAXI~J1807+132 during the decay of its 2017 outburst. We carried out a photometric follow-up in three SDSS bands (\textit{g}, \textit{r}, and \textit{i}). Simultaneous optical spectroscopy was performed in 6 different epochs, both in outburst and quiescence. In addition, we analyse \textit{XMM-Newton} archival observations taken soon after the MAXI alert.

\section{Observations and data reduction}

\subsection{Optical data}
The photometric observations were carried out from March 28 to July 12 of 2017 over 35 different epochs. The data were obtained in the SDSS-\textit{g},-\textit{r}, and -\textit{i} bands using the 2m Liverpool Telescope (LT) at the Observatorio del Roque de Los Muchachos (hereafter ORM) located in La Palma (Spain), and the 2-meter and 1-meter class telescopes from Las Cumbres Observatory (LCO). In addition, a continuous light-curve of 82 photometric points was taken (SDSS-\textit{r}) using the 4.2-m William Herschel Telescope (WHT) at the ORM over a time-lapse of $\sim 160$ minutes on July 22. The WHT data were reduced using \textsc{astropy-ccdproc} based routines \citep{Astropy2013}. Data from LCO were automatically processed through the \textsc{banzai} pipeline\footnote{Beautiful Algorithms to Normalize Zillions of Astronomical Images. Code available at https://github.com/LCOGT/banzai}. The LT data reduction was performed using the IO:O data reduction pipeline\footnote{http://telescope.livjm.ac.uk/TelInst/Pipelines/\#ioo}. Flux calibration was carried out against nearby stars present in the PanSTARRs catalog.

We also obtained intermediate resolution spectroscopy using the Optical System for Imaging and low-Intermediate Resolution Integrated Spectroscopy \citep[OSIRIS;][]{Cepa2000} attached to 10.4m Gran Telescopio Canarias (GTC) at the ORM. We used the R1000B optical grism (2.12~\AA\ pix$^{-1}$ at 5455~\AA) covering the spectral range 3630--7500~\AA. This, in combination with a slit width of 1 arcsec, provided a spectral resolution of 360~km~s$^{-1}$ (measured as the full-width at half-maximum at $\sim$ 5577~\AA). A total of 5 spectra were obtained from March 28 to August 18. In addition, one spectrum was taken on July 2018 using the R1000R grism (2.62~\AA\ pix$^{-1}$ at 7430~\AA) covering the range 5100--10000~\AA. We reached a resolution of 381~km~s$^{-1}$ (full-width at half-maximum at $\sim$ 5577~\AA) using a slit width of 1 arcsec.

The spectroscopic data were de-biased and flat-fielded using \textsc{iraf} standard routines. We used regular arc lamp exposures taken on each observing night to carry out the pixel-to-wavelength calibration (HgAr+Ne lamps for the R1000B grism and HgAr+Ne+Xe for the R1000R). Cosmic rays were removed from the data using L.A.Cosmic \citep{Dokkum2001} before extracting the spectra. We used the \ion{O}{i} 5577.338~\AA\ sky emission line and the \textsc{molly} software to measure and then correct the subpixel velocity drifts (<~100~km~s$^{-1}$) introduced by instrumental flexure effects.
 
In addition, we used the acquisition images (SDSS-\textit{g} and SDSS-\textit{r}) to obtain photometric measurements contemporaneous with the spectroscopic observations (see Table \ref{tab:journald_spect}). These were reduced in the same way as the WHT observations.

\begin{table}
	\centering
	\caption{Spectroscopy of MAXI~J1807+132.}
	\label{tab:journald_spect}
	\begin{tabular}{lccr} 
		\hline
		Spectrum  & Date & Exp. time (s) & SDSS-\textit{g} (mag) \\
		\hline
		1 & 28/03/2017 & 600 (2 $\times$ 300) & $18.36\pm0.01$\\
		2 & 30/03/2017 & 900 (2 $\times$ 450) & $19.16 \pm0.03$\\
		3 & 06/04/2017 & 3600 (2 $\times$ 1800) & $19.08 \pm0.02$\\
		4 & 16/07/2017 & 3000 (2 $\times$ 1500) &$21.60\pm0.23$\\
		5 & 18/08/2017 & 1003 & $21.51\pm0.08$\\
		6 & 13/07/2018 & 1800 & $21.36\pm0.01*$\\
		\hline
		*SDSS-\textit{r}		

	\end{tabular}
\end{table}

\subsection{X-ray observations}\label{subsec:X-ray}
\subsubsection{\textit{Swift} data}\label{subsec:swift data}

The Neil Gehrels \swift\ Observatory \citep{Gehrels2004} pointed to MAXI~J1807+132 on 29 occasions since it was reported as an active source. The on-board X-ray Telescope (XRT; \citealt{Burrows2005}) was operated in window timing (WT) mode for the first 4 pointings and in photon counting (PC) mode for the remaining observations. We used the {\ttfamily HEASoft} v.6.20 package to reduce the data. We processed the raw XRT data running the {\ttfamily xrtpipeline} task selecting the standard event grades of 0-12 and 0-2 for the PC and WT mode observations, respectively. With {\ttfamily Xselect}~v2.4 we extracted the source events from a circular region of $\sim40$~arcsec radius. To compute the background, we used three circular regions of similar size and shape on an empty sky region (PC observations), and an annulus centred on the source with an 195~arcsec inner radius and 275~arcsec outer radius for the WT data. 

The source is detected in 16 out of the 29 observations (i.e., with signal-to-noise ratio > 3~$\sigma$), and only 4 of them have enough counts to carry out spectral analysis using \chis\ statistics (> 50 counts; see Fig. \ref{fig:allphot} top panel). For these observations, we created exposure maps and ancillary response files following the standard \textit{Swift} analysis threads\footnote{\url{http://www.swift.ac.uk/analysis/xrt/}}, and we acquired the last version of the response matrix files from the High Energy Astrophysics Science Archive Research Center (\textsc{HEASARC}) calibration database (CALDB). We grouped the spectra so as to attain a minimum of 10 photons per bin and, therefore, be able to use \chis\ statistics consistently.

\subsubsection{\textit{XMM-Newton} data}

Two observations were acquired with the \xmm\ observatory \citep{Jansen2001} on March 29 and 30, 2017 (gap between them of less than 7~h), with exposures of 19 and 29~ksec, respectively. The configuration of the European Photon Imaging Camera (EPIC) was the same for both observations: MOS1 detector was operated in imaging small-window mode, and MOS2 and PN detectors in timing mode, all of them with the thin filter \citep{Turner2001,Strueder2001}. The source was not detected in the MOS2 camera.
Due to strong episodes of background flaring, we proceeded to exclude data with count rates $>$~0.4~$\cnts$ at energies $>$10~keV and $>$~0.5~$\cnts$ at energies 10--20~keV for the MOS and PN cameras, respectively. 
For MOS1, we extracted source events from a circular region with a radius of 45~arcsec and a source-free circular region with a radius of 100~arcsec for the background. For PN, we extracted source and background events using the RAWX columns in [33:43] and in [10:18], respectively. Light curves and spectra, as well as associated response matrix files and ancillary response files were generated following the standard analysis threads\footnote{\url{https://www.cosmos.esa.int/web/xmm-newton/sas-threads}}. For MOS1, we rebinned the spectrum in order to include a minimum of 25 counts in every spectral channel, avoiding to oversample the full width at half-maximum of the energy resolution by a factor larger than 3. In the case of PN, we rebinned the spectrum in order to have a minimum signal-to-noise ratio of 6.

%
\begin{table*}
\caption{Results from the X-ray spectral fits.}
\begin{threeparttable}
\begin{tabular}{l c c c c c c  }
\hline \hline
Instrument/ &  $kT_{\rm bb/in}$ & $N_{\rm bb/in}$ & $\Gamma$  &  $F_{\mathrm{X,unabs}}$ & $T_{\mathrm{fr}}$   & $\chi^2_{\nu}$ (dof) \\
Obs ID 	    & (keV) &              &            & ($10^{-12}$ $\flux$)   &		(\%)    			&  					\\  
\hline
\swift/XRT & \multicolumn{6}{c}{\textsc{Tbabs*(powerlaw)}}\\

00010037001  & -- &  -- & $2.01 \pm 0.06$ & $19.1 \pm 1.5$    &   --  	  &1.37 (145)\\ 
00010037002  & -- &  -- & $2.1 \pm 0.1$   & $5.6  \pm 1.0$    &   --        & 0.89 (59)\\ 
00010037004  & -- &  -- & $2.6 \pm 0.8$   & $0.33 \pm 0.1$       &   --        & 0.62 (25)\\ 
00010037017  & -- &  -- & $1.8 \pm 0.2$   & $5.7 \pm 1.4$    &   --        & 1.06 (22)\\ 
\hline 
\textit{XMM}/EPIC & \multicolumn{6}{c}{\textsc{Tbabs*(bbodyrad+nthcomp(bb))}}\\

		  &  $0.21 \pm 0.01$ &  $111 \pm 10 $ & $2.1 \pm 0.1$  &  $6.42 \pm 0.1 $& 25 & 1.01 (576) \\

 		  & \multicolumn{6}{c}{\textsc{Tbabs*(diskbb+nthcomp(disk))}}\\

	      &  $0.33 \pm 0.01$ &  $15 \pm 2 $ & $1.8 \pm 0.1$ & 6.7 $\pm$ 0.1 & 36  & 0.96 (576) \\

\hline
\end{tabular}
\begin{tablenotes}
\item[]Note. -- Quoted errors represent 90\% confidence levels. $F_{\mathrm{X, unabs}}$ represent the unabsorbed fluxes in the 2--10~keV and 0.5--10~keV bands for \swift\ and \xmm\ observations, respectively. $T_{\mathrm{fr}}$ reflects the fractional contribution of the thermal component to the total unabsorbed 0.5--10 keV flux. When using NTHCOMP we coupled $kT_{\rm seed}$ to the temperature of the thermal component (i.e., $kT_{\rm disc}$ or $kT_{\rm bb}$; see Section \ref{subsub:xmm})
\end{tablenotes}
\label{tab:results}
\end{threeparttable}
\end{table*}


\section{Results}

\subsection{Outburst evolution}\label{outburst_evolution}

\begin{figure}
	\includegraphics[width=\columnwidth]{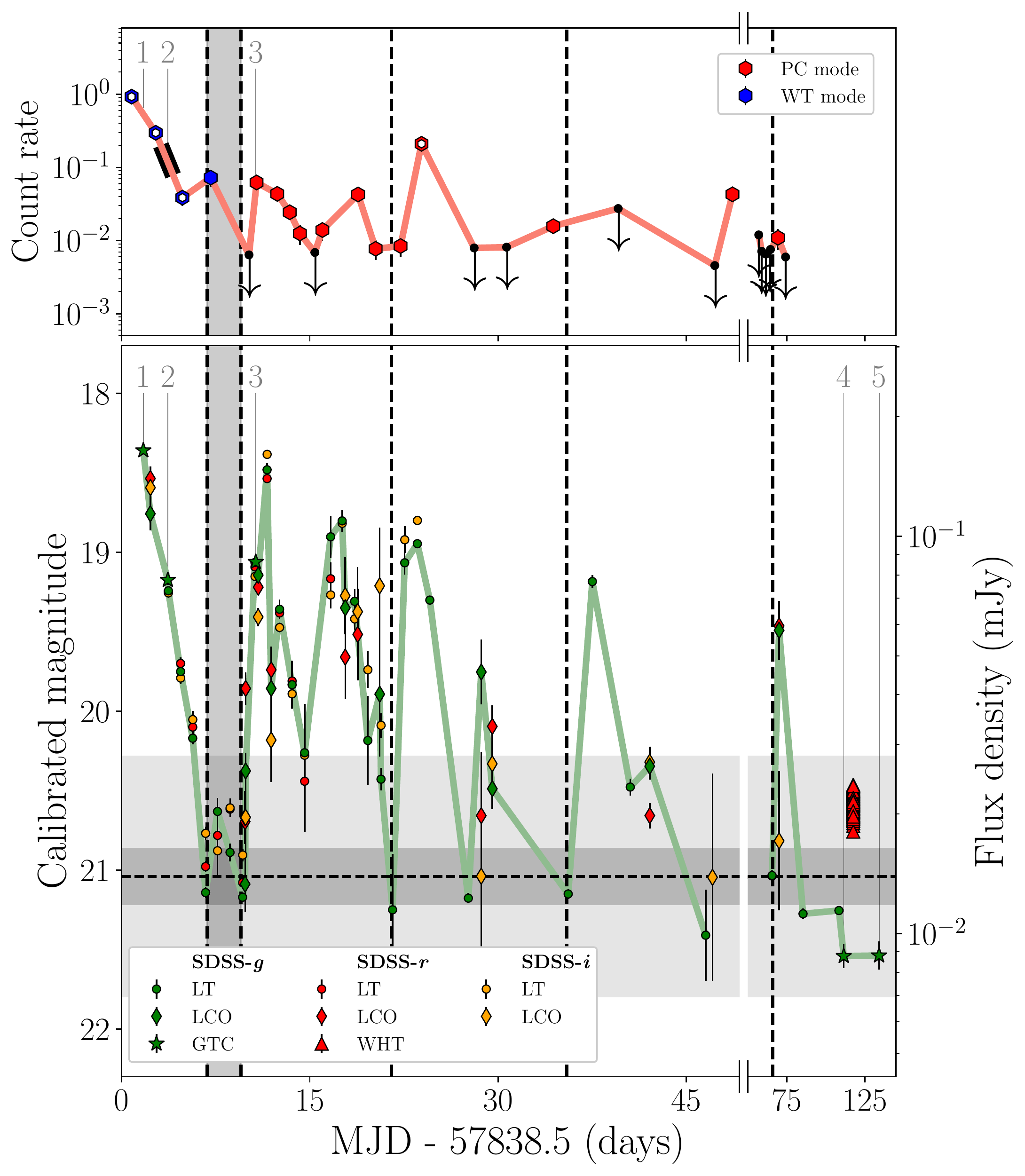}
	\includegraphics[width=\columnwidth]{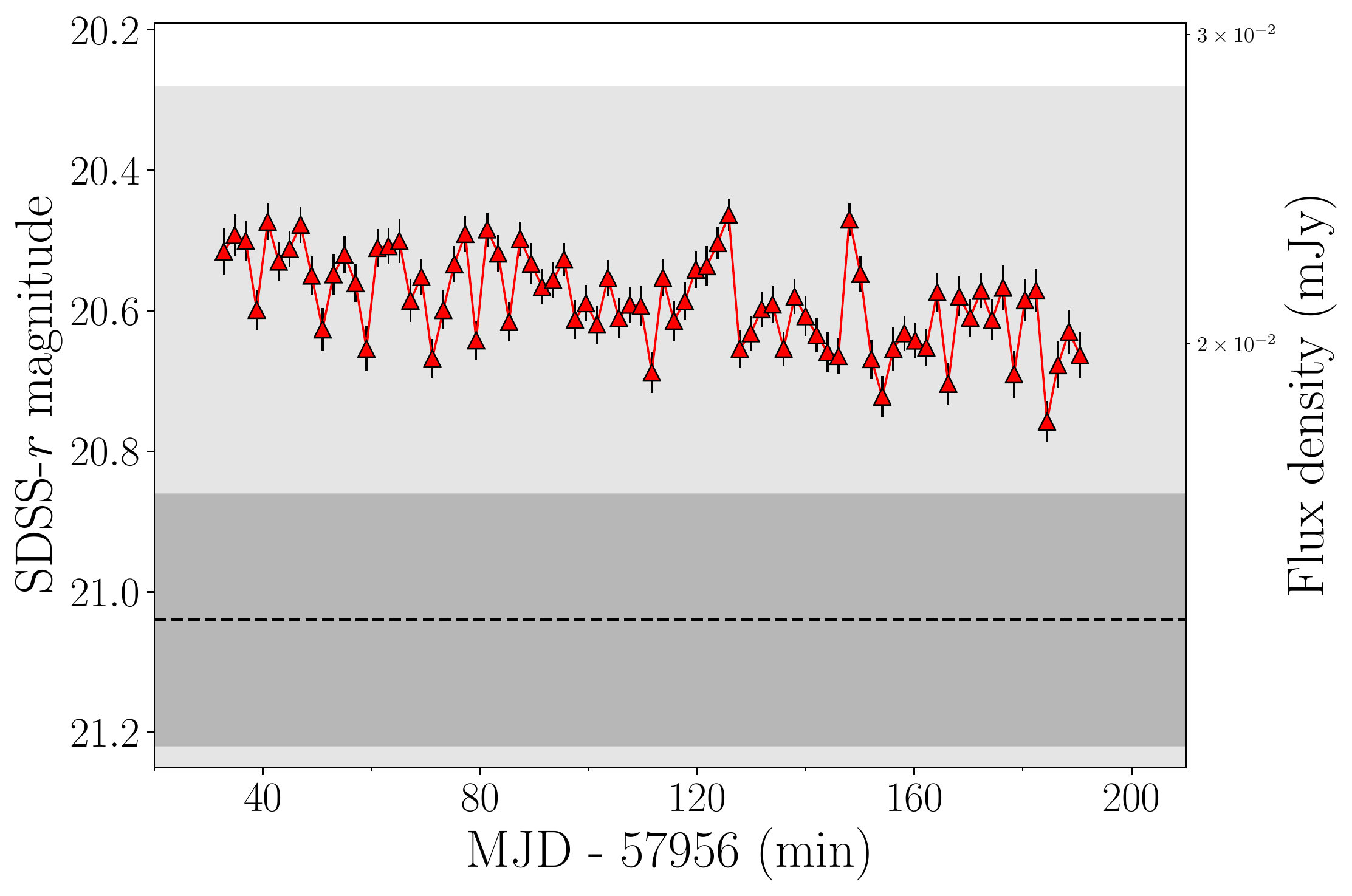}

    \caption{Top panel: X-ray light curve obtained with \swift/XRT. The thick black line indicates the \xmm\ epochs. The observations employed in the spectral analysis are indicated as open symbols. Middle panel: Photometric follow-up of MAXI~J1807+132 in the SDSS-\textit{g}, -\textit{r}, and -\textit{i} bands (green, red and yellow markers, respectively). LT data are indicated by dots, while LCO data are marked as diamonds. At the end of the series, triangles show WHT points. GTC photometry from acquisition images is represented as stars and labelled in correspondence with spectra, except for the SDSS-\textit{r} point and its corresponding spectrum taken on July 2018 ; Table \ref{tab:journald_spect}. The light green, solid line joins the \textit{g}-band points chronologically. The black dashed line shows the quiescence PanSTARRS \textit{g}-band magnitude (after stacking) with its corresponding error displayed as dark grey area \citep[$21.04\pm0.18$;][]{Denisenko2017}. The standard deviation of the 9 pre-outburst epochs is indicated as a light grey area. The vertical dashed lines (and grey band) indicate epochs consistent with optical quiescence where the source is detected in X-rays. Bottom panel: Zoom in of the WHT light-curve spanning 160 min.}
    \label{fig:allphot}
\end{figure}

Our photometric follow-up is presented in the middle panel of Fig.~\ref{fig:allphot} and Table \ref{tab:journald_phot}. It mostly includes photometric points in the SDSS-\textit{g} band complemented with measurements in SDSS-\textit{r} and -\textit{i}. The photometric monitoring started while the source was decaying from the outburst peak. Our first data point is almost 3 mag brighter than the quiescence magnitude reported by \cite{Denisenko2017}. Over the next 9 days the source gradually decayed from \textit{g} $\sim18.8$ to quiescence (\textit{g} $\sim21$). Subsequently, we observed an abrupt rise of $\sim2.7$ mag in two days. We identify up to 7 re-brightening events superposed on a decay profile during the $\sim70$ days following the main outburst. This activity is detected in all the photometric bands studied. The secondary peaks have amplitudes of $\sim 2$ mag above the quiescent level. A tentative periodic recurrence of $ \sim 6$ days is observed in the brightest 4 events (see Section \ref{reflares}). 

\begin{figure*}
	\includegraphics[width=\textwidth]{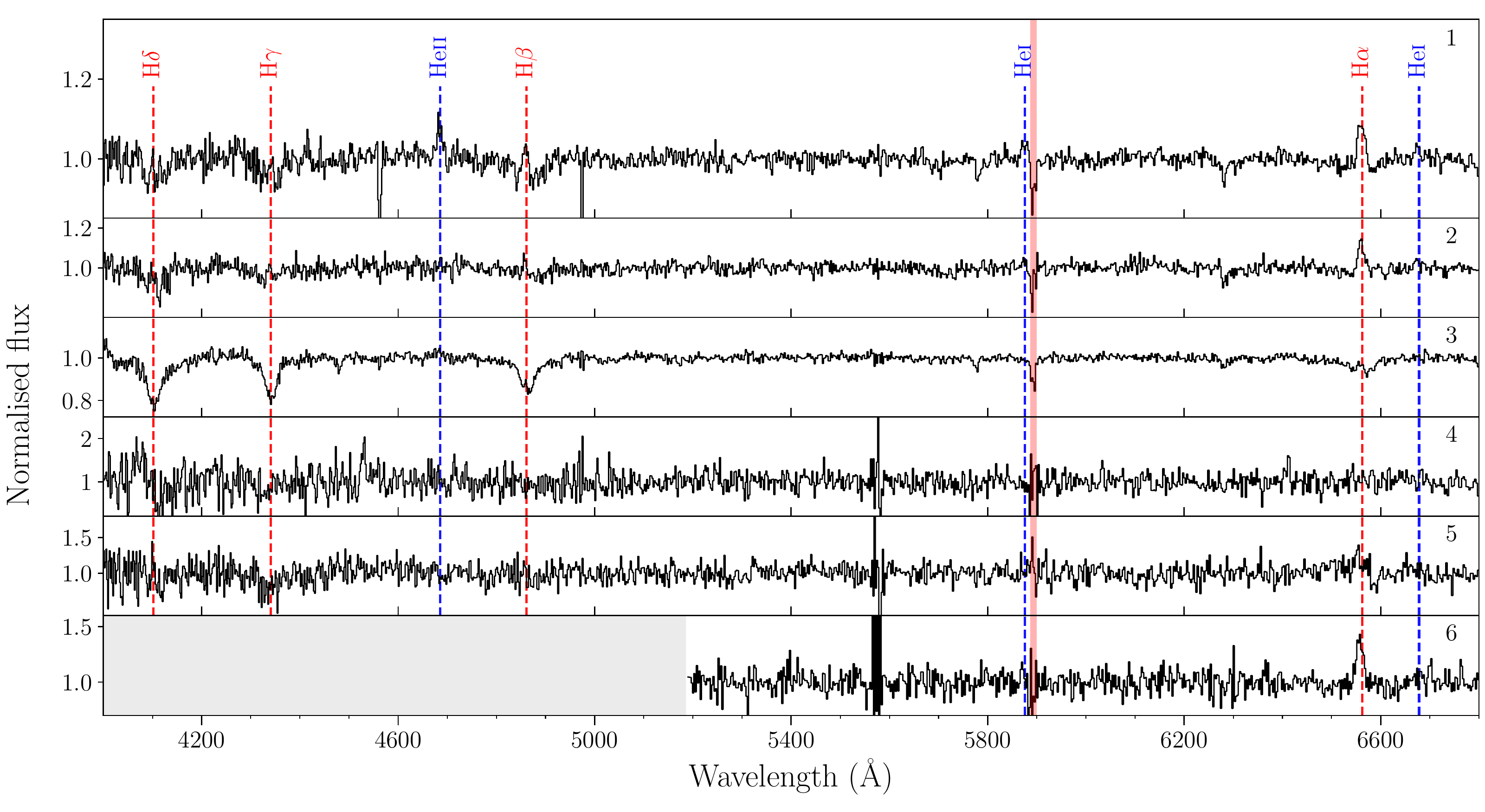}
    \caption{Multi-epoch spectroscopy of MAXI~J1807+132 from GTC-OSIRIS (R1000B grism spectrum-1 to -5 and R1000R spectrum-6). Hydrogen (red) and He (blue) transitions are indicated by dashed lines. The \ion{Na}{i} doublet is indicated as a red shadowed area at 5892~\AA. Spectrum-4, -5 , and -6 show strong residuals caused by sky subtraction at $\sim$ 5577 and $\sim$ 5890~\AA.}

    \label{fig:all_spec}
\end{figure*}

During the photometric monitoring, 6 optical spectra were obtained (and labelled chronologically spectrum-1 to -6 hereafter). These are presented in Fig. \ref{fig:all_spec}. The first 3 spectra were taken when the source was brighter than \textit{g} $\sim$ 19.5, while the remaining found the source in quiescence at \textit{g} $\sim21$ (see Fig. \ref{fig:allphot} and Table \ref{tab:journald_spect}). Hereafter, we refer to them as bright and faint spectra, respectively. The source showed pronounced spectral variability as it evolves through its complex outburst decay. We observed several emission lines in the bright spectra that are commonly seen in LMXBs in outburst. For example, spectrum-1 exhibits the Balmer series up to H$\gamma$ as well as \ion{He}{II} 4686~\AA. Less evident \ion{He}{i} lines at $\sim$ 5875~\AA\ and $\sim6679$~\AA\ can be also identified. Fainter Balmer lines are present in the remaining bright spectra. Conversely, we can not distinguish He transitions in spectrum-2 and -3. In addition, the Balmer emission lines appear embedded in broad absorptions that are stronger for the bluest lines. This pattern is particularly noticeable in spectrum-3, but it is still recognizable in spectrum-1 and -2. We note that spectrum-2 and -3 were obtained at comparable magnitudes ($19.22\pm0.02$ and $19.06 \pm0.02$ SDSS-\textit{g} mag, respectively) but they show significant differences. Spectrum-2 is morphologically closer to spectrum-1. However, for spectrum-3 the broad absorptions are remarkably deep and dominate over the emission lines.

We also observed morphological differences between faint spectra. Spectrum-4 and -5 are featureless while spectrum-6 only exhibits  H$\alpha$ in emission.  Absorption lines that could be associated with the companion star are not detected in any of the spectra.

\subsection{Emission lines properties}
\label{lines}

\begin{figure}
	\includegraphics[width=\columnwidth]{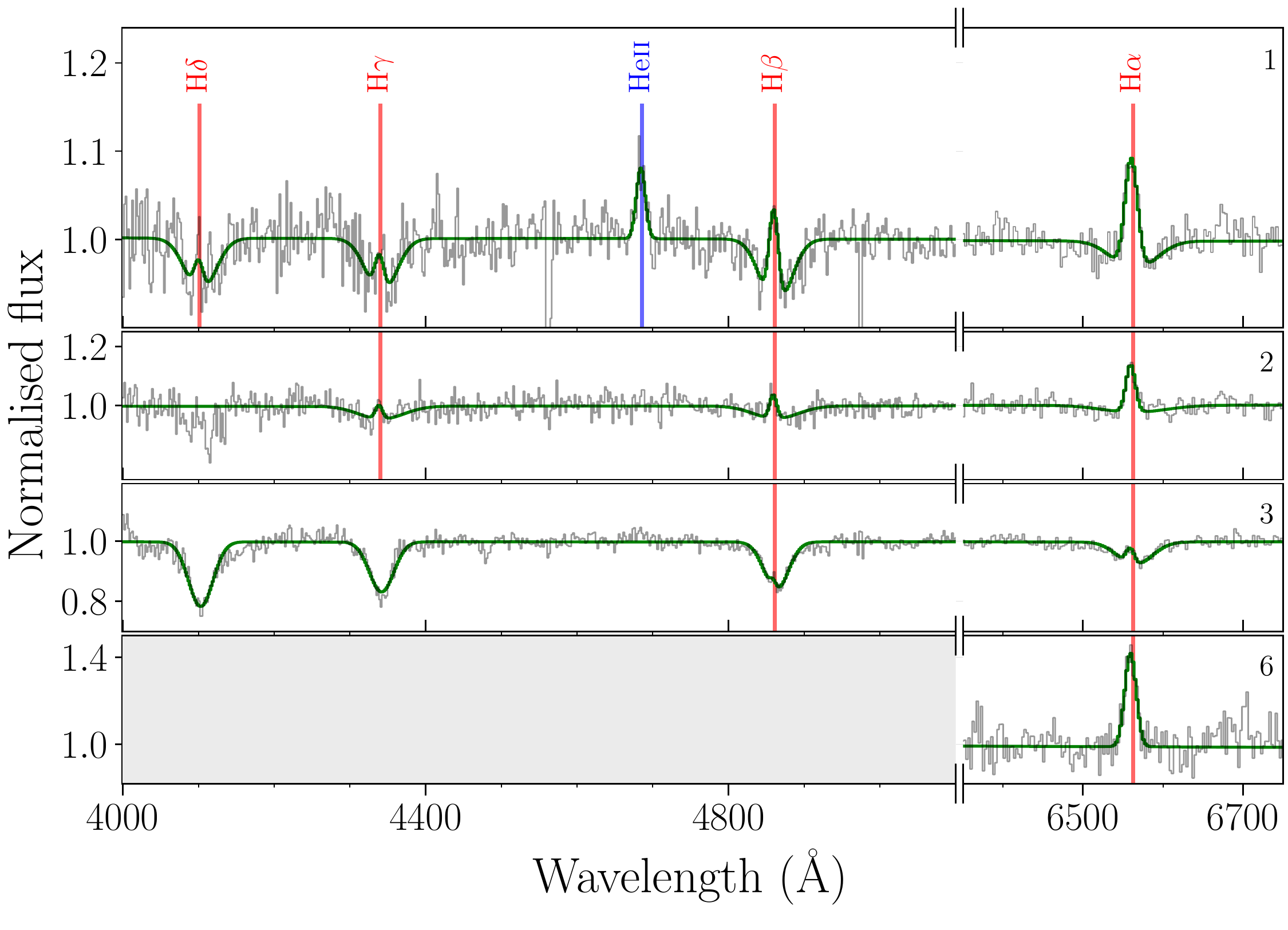}
    \caption{Trimmed spectra showing identified emission lines. The vertical lines represent the rest frame position of the transition used on each model (red for the Balmer series and blue for \ion{He}{ii} 4686~\AA). Green line shows the resulting model.}
    \label{fig:4_spect_1_3}
\end{figure}

We measured Doppler shifts in the emission lines observed in the spectrum-1,-2, -3, and -6 by fitting a multi-Gaussian model to previously normalised spectra. The lines considered and the model applied to each spectrum varies slightly from one to another following the morphological differences between them. For instance we modelled the Balmer lines and \ion{He}{ii} emission line in spectrum-1, while only  H$\alpha$ was fitted in spectrum-6. In order to obtain more reliable velocity measurements the separations between Gaussians were fixed. The broad absorptions underneath were taken into account when modelling the Balmer lines in spectrum-1,-2, and -3. To this end, we fitted broad Gaussian profiles keeping the relative separations between them fixed as in the case of the emission lines. We used a 2-Gaussian model to account for the broad absorption components in the line profiles when present. The broadening effect on the emission lines due to the instrumental resolution was taken into account in the modelling. The results are shown in Fig. \ref{fig:4_spect_1_3}.

The derived emission line centroid velocities are presented in Table \ref{tab:parameter}. In spectrum-3 the velocity obtained relies only on 2 emission lines strongly affected by the absorptions, and thus the uncertainty on the centroid velocity is a factor of $\sim 2$ larger than that from spectrum-1. The derived velocities are consistent with each other within 1.5 $\sigma$. We can tentatively associate these velocity shifts with the velocity of the centre of mass of the system, projected along the line of sight, $\gamma$. We note that, on one hand, it seems unlikely that the 3 outburst spectra were taken at the same orbital phase and reflect the projected velocity of a particular structure of the accretion disc. On the other hand, it is highly unlikely that these velocities arise from some region (or regions) in the system showing the same radial velocity at different orbital phases. This argument is strongly favoured by the velocity measured from spectrum-6, taken more than a year later. We propose the average value $\gamma=-145\pm13$~km~s$^{-1}$ as the systemic velocity of the binary, although a radial velocity curve is required to obtain a conclusive result.

The offset measured in the absorption components observed in spectrum-1, -2, and -3 are centred at $\sim80$~km~s$^{-1}$ showing uncertainties bigger than $\sim 190$~km~s$^{-1}$. Hence, we cannot constrain the velocity from these offsets.
  
We determined the FWHM of the emission lines from the multi-Gaussian fitting described above. In addition, we individually fitted the  H$\alpha$ emission line and its broad absorption profile with a double Gaussian model in spectrum-1,-2, and -3. Table \ref{tab:parameter} shows both the FWHM from all the emission lines considered and that from  H$\alpha$ for each spectra. The measured FWHM of the broad components are $2851\pm223$, $4260\pm531$, and $2303\pm45$~km~s$^{-1}$ for spectrum-1,-2, and 3, respectively.
Finally, the equivalent width (EW) of the H$\alpha$ emission line in the spectra which are not contaminated by the absorption component are $1.96\pm0.11$~\AA\ (spectrum-1), $2.89\pm0.20$~\AA\ (spectrum-2) and $6.86\pm0.70$~\AA\ (spectrum-6), whilst \ion{He}{ii} has EW $1.35\pm0.17$~\AA\ in spectrum-1.

\begin{table}
	\centering
	\caption{ Parameters from the observed emission lines. FWHM refers to the ensemble of all the emission lines}
	\label{tab:parameter}
	\begin{tabular}{lccc} 
		\hline
		Spectrum  & FWHM  & Centroid velocity & H$\alpha$ FWHM  \\
		  &(km~s$^{-1}$) & (km~s$^{-1}$) &(km~s$^{-1}$) \\

		\hline
		1 & $796\pm57$ & $-106\pm17$ &$750\pm63$\\
		2 & $557\pm64$ & $-147\pm21$&$630\pm74$\\
		3 & $468\pm106$& $-169\pm31$&$499\pm142$\\
		6 & --&$-157\pm35$&$698\pm89$\\
		\hline
	\end{tabular}
\end{table}

\subsection{Extinction along the line of sight}\label{Extinction}

The depth of interstellar absorption bands can be used to estimate the extinction along the line of sight. In particular, the color excess E$_{\mathrm{B}-\mathrm{V}}$ is correlated with the EW of the strongest component of the \ion{Na}{i} doublet (NaD1, 5890~\AA). This correlation is valid for EWs $\lesssim$~0.5~\AA\ \citep{Munari1997}. In the bright spectra we observed the two components of the \ion{Na}{i} doublet blended in a single absorption feature. No \ion{Na}{i} components were detected in the faint spectra, which are contaminated by nearby sky emission lines (see Fig. \ref{fig:all_spec}). We averaged the 3 bright spectra and derived EW~$=-1.10\pm0.08$~\AA. This implies that the line is saturated and only a lower limit E$_{\mathrm{B}-\mathrm{V}}\geq0.3$ can be derived. This corresponds to N$_{\mathrm{H}}\gtrapprox2.4\times10^{21}$cm$^{-2}$ when applying the dust to gas relation \citep[N$_{\mathrm{H}}=2.87\times10^{21}$cm$^{-2}$~A$_{\mathrm{V}}$;][]{Foight2016} converting using the canonical relation A$_{\mathrm{V}}=3$E$_{\mathrm{B}-\mathrm{V}}$. This value is still roughly consistent with the Galactic value, N$_{\mathrm{H}}\sim1\times10^{21}$cm$^{-2}$ \citep{Kalberla2005}.

\subsection{X-ray analysis}
We used \textsc{xspec} (v.12.9.1; \citealt{Arnaud1996}) to analyse the X-ray spectra. In order to account for interstellar absorption, we used the Tuebingen-Boulder Interstellar Medium absorption model (TBABS in \textsc{xspec}) with cross-sections of \citet{Verner1996} and abundances of \citet{Wilms2000}. We assumed a constant hydrogen equivalent column density of $N_{\rm  H}=1\times10^{21}~\rm cm^{-2}$ consistent with the low reddening expected for this direction (see previous section) and \citet{Shidatsu2017}.
\subsubsection{\swift}
We used a simple power-law model to fit the 4 \swift\ spectra. We calculated the 2--10~keV unabsorbed flux in order to plot these points in the Optical/X-ray correlation (see Section \ref{OpticalXray}). We refer the reader to \citet{Shidatsu2017} for more details on the \swift\ observations. 

\subsubsection{\xmm}\label{subsub:xmm}
We simultaneously fit the 0.5-10~keV MOS1 and 0.7-10~keV PN spectra by using tied spectral parameters of both \xmm\ observations. We included a constant factor (CONSTANT) fixed to 1 for PN spectra and free to vary for MOS1 spectra so to account for cross-calibration uncertainties between the instruments.

\begin{figure}
	\includegraphics[width=\columnwidth]{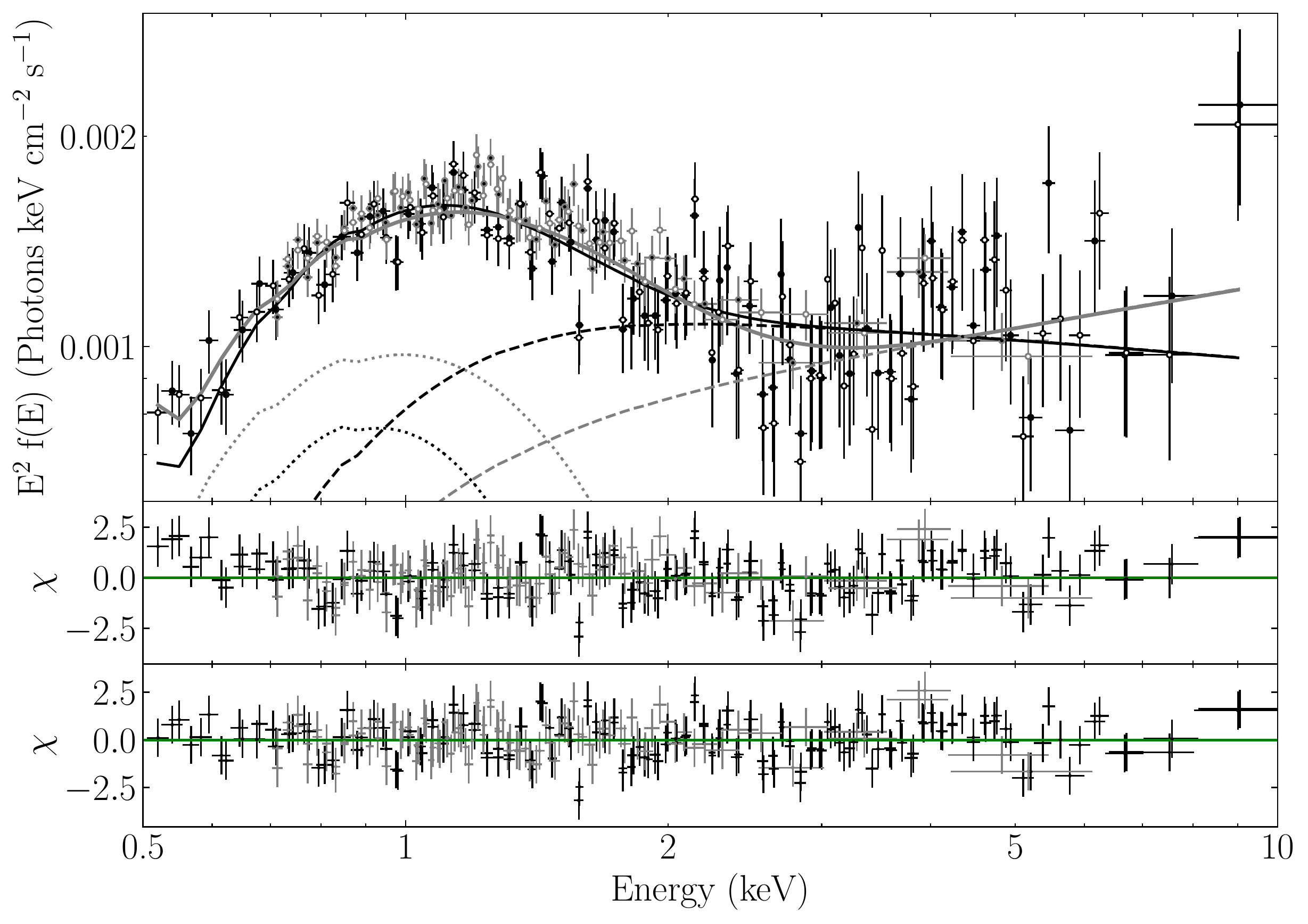}
    \caption{ Top panel: Unfolded MOS1 (black) and PN (grey) spectra of the March 29 (filled symbol) and 30 (open symbol) \xmm\ observations. The solid black line represents the best fit for the BBODYRAD+NTHCOMP model, the thermal component is shown as a dotted line and the Comptonized component as a dashed line. In grey is represented the same for the DISKBB+NTHCOMP model. Middle panel: Residuals in units of $\sigma$ when using BBODYRAD+NTHCOMP model. Botton panel: Residuals in units of $\sigma$ when using DISKBB+NTHCOMP model.}
	\label{fig:spectra}    
\end{figure}

We used a simple 2-component model to fit the spectra. We combined a soft component to account for either emission from an accretion disc or from a possible NS surface/boundary-layer, and a hard component to model the inverse-Compton emission from the Corona. Thus, we used a multicolour disc or single black body (BB) plus a thermally Comptonized continuum model (i.e., DISKBB+NTHCOMP and BBODYRAD+NTHCOMP in \textsc{xspec}; \citealt{Mitsuda1984,Makishima1986, Zdziarski1996, Zycki1999}). We assumed that the up-scattered photons arise from the associated thermal component. Thus, we coupled $kT_{\rm seed}$ to either $kT_{\rm disc}$ or $kT_{\rm bb}$, respectively, and changed the seed photons shape parameter accordingly (\textit{im-type} in NTHCOMP). The two spectral models return acceptable fits (\chis\ $\sim$ 544-584 for 576 dof). The former yields a Comptonization asymptotic power-law photon index ($\Gamma$) of $1.8\pm0.1$ and a disc temperature of $kT_{\rm diskbb}=0.33\pm0.01$~keV with a $N_{\rm in}=15\pm2$. The latter translates to an unrealistic small internal disc radius ($R_{\rm in}\sim5$~km applying the correction of \citealt{Kubota1998}; see also \citealt{ArmasPadilla2017b}, and references therein) and compact object mass for a BH case \citep[e.g.,][]{Tomsick2009}. However, we note that it is not straightforward to obtain physically meaningful  $R_{\rm in}$ values from this kind of modelling. We further discuss this in section \ref{co}. The inferred 0.5--10~keV unabsorbed flux is ($6.7\pm0.1)\times10^{-12}~\flux$, from which $\sim$ 36~per cent is produced by the thermal component. On the other hand, by using BBODYRAD+NTHCOMP we obtained $kT_{\rm bb}=0.21\pm0.01$~keV and $N_{\rm bb}=111\pm10$ while $\Gamma$ is $2.1\pm0.1$. The inferred 0.5--10~keV unabsorbed flux is ($6.4\pm0.2)\times10^{-12}~\flux$, from which $\sim$ 25~per cent is produced by the thermal component. It is important to bear in mind that due to the lack of data above 10~keV and the low statistics, we were not able to constrain the temperature of the electron corona ($kT_{\rm e}$), that was fixed to either 20~keV or 100~keV, typical values for NSs and BHs in the hard state \citep{Burke2017}. This value do not have an impact on the obtained results (within errors). The best fit results are reported in Table \ref{tab:results} and Fig. \ref{fig:spectra} with uncertainties given at 90 per cent confidence level. 

As described above, the 2-component approach provides a good description of the data. However, we also tried other solutions. For instance, a simpler one-component model using a NTHCOMP was tested considering both scenarios: seed photons arising from a blackbody or from a disc blackbody. As expected, this forced the photon index to adopt higher values ($\Gamma=2.48\pm0.03$) leaving significant residuals at high energies (\chis\ of 652 and 665 for 577 dof, respectively). This is reflected in the F-test, which yields a probability lower than $10^{-15}$ of improving the fit by chance when including a thermal component. In addition, more complex X-ray fitting, such as that provided by the three-component model proposed for NSs \citep[e.g.,][]{Lin2007,ArmasPadilla2017b} was also considered. However, even if this also reproduces the data, it does not produces a significant improvement and its use is not justified given the limited quality of the data and the relatively narrow spectral coverage of \xmm.

\subsection{Optical/X-ray correlation}\label{OpticalXray}

Based on quasi-simultaneous optical/near-infrarred and X-ray observations of a large number of BHs and NSs, \cite{Russell2006} presented an empirical correlation between the nature of the compact object and its optical to X-ray luminosity ratio.  

\begin{figure}
	\includegraphics[width=\columnwidth]{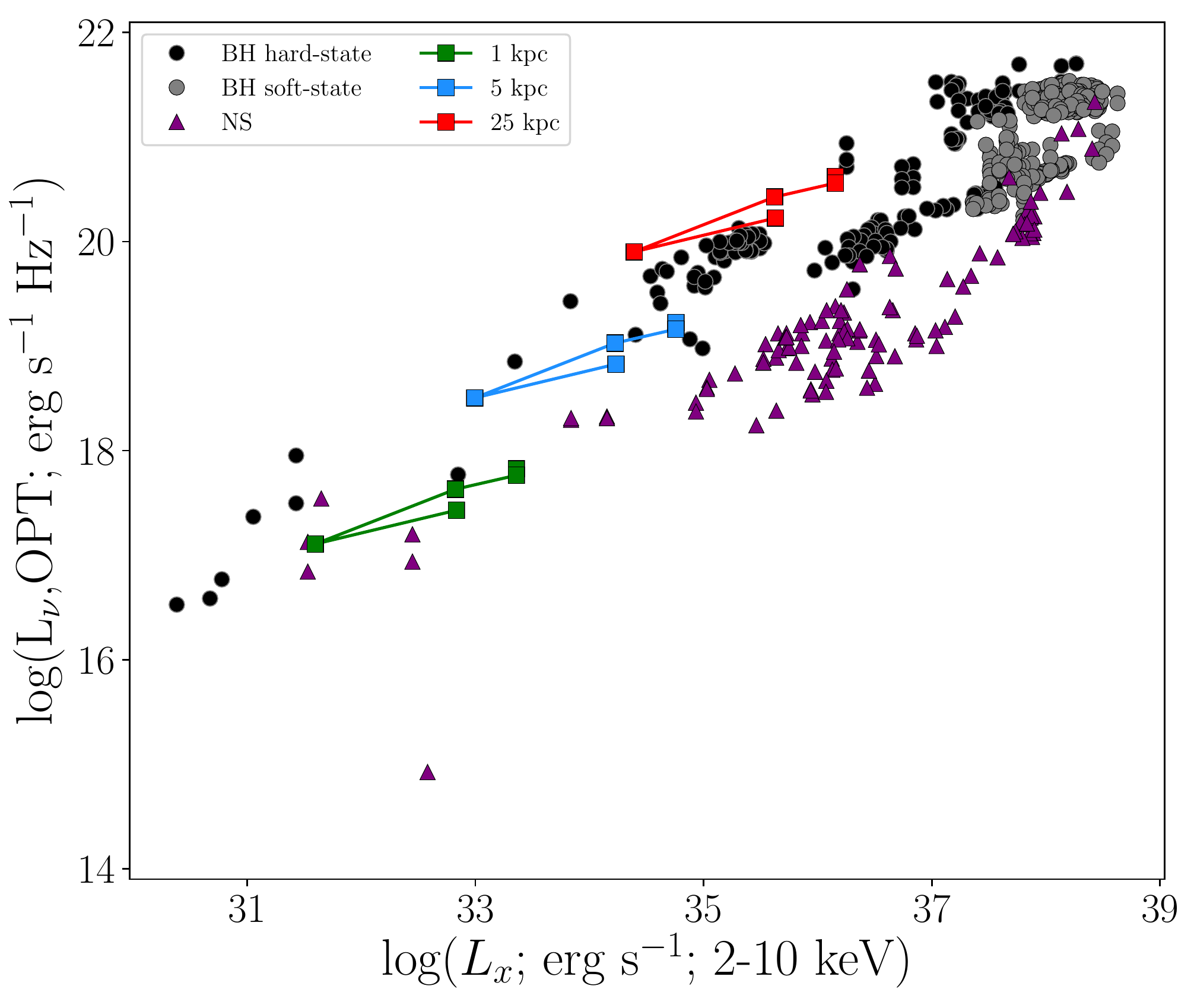}
    \caption{Optical-X-ray luminosity diagram \citep[data from][]{Russell2006,Russell2007}. BHs in the hard and soft states are indicated as black and grey circles, respectively; NSs are plotted as violet triangles. Our data for MAXIJ1807+132 are overplotted assuming 3 distances: 1~kpc (green square), 5~kpc (blue square), and 25~kpc (red square).}
    \label{fig:OptXrayCorr}
\end{figure}

Fig. \ref{fig:OptXrayCorr} shows the optical-X-ray luminosity diagram plotted using data from \cite{Russell2006,Russell2007}. We investigate the location of MAXI~J1807+132 on this diagram in order to constrain the nature of the compact object. To this end, we use 5 luminosity ratios, obtained from a quasi-simultaneous optical and X-ray data. To compute the luminosities, we consider 3 possibles distances: 1, 5, and 25~kpc. The optical/X-ray luminosity ratio suggests a BH nature for distances $\gtrsim 5$~kpc. For closer distances the source sits on a transitional and poorly sampled region in between the BH and NS populations. A NS accretor is favoured if the distance is $\sim$ 1~kpc or lower.

\subsection{Quiescence photometry}

The bottom panel of Fig. \ref{fig:allphot} shows the SDSS-\textit{r} light curve of MAXIJ1807+132 obtained on July 22, 2017 using the WHT. The source showed a mean brightness of $20.584\pm0.003$ mag, that is, $\sim0.6$ mag above the quiescence level reporter by \cite{Denisenko2017}. The light curve shows a complex behaviour dominated by flickering with an amplitude of up to $\sim 0.3$ mag and a variance of $0.06$ mag with respect to the mean (this correspond to 0.0013 mJy over a mean of 0.021 mJy). We explored the WHT dataset seeking for periodic oscillations using standard Lomb-Scargle techniques, but no modulations were found within the $\sim2.5$ hours observed.

\section{Discussion}

We presented a detailed optical and X-ray follow up of MAXI~J1807+132 during its 2017 outburst decay. This includes optical spectroscopy using GTC-10.4m, which reveals emission lines typically observed in LMXB spectra. The Balmer series appears in emission superposed to broad absorption features that are deeper towards the blue. Similar line profiles were observed before in other LMXBs \citep[e.g., Nova~Velorum~1993, GRO~J0422+32, XTE~J1118+480, XTE~J1859+226;][respectively]{Bailyn1995,Casares1995a, Dubus2001,Zurita2002}. This can be interpreted as the result of emission lines being originated from photoionization in an optically thin region above the disc, combined with absorption lines arising from inner optically thick regions. The absorption throats were discussed in detail by \citet{Dubus2001}. They found that the formation of these features is favoured by (hard X-ray) irradiation, since hard X-ray photons can access deeper disc. In addition, these features are not expected in high inclination systems, since the absorptions tend to disappear as the inclination increases, affected by the limb darkening in the two-dimensional disc picture. If this interpretation is correct, it would suggest an intermediate to low orbital inclination for the system. However, we note that both GRO~J0422+32 and XTE J1118+480 have intermediate to high inclinations \citep[45$^{\circ}\pm 2$ and 68--79$^{\circ}$;][]{Gelino2003,Khargharia2012}, respectively.

\begin{figure}
	\includegraphics[width=\columnwidth]{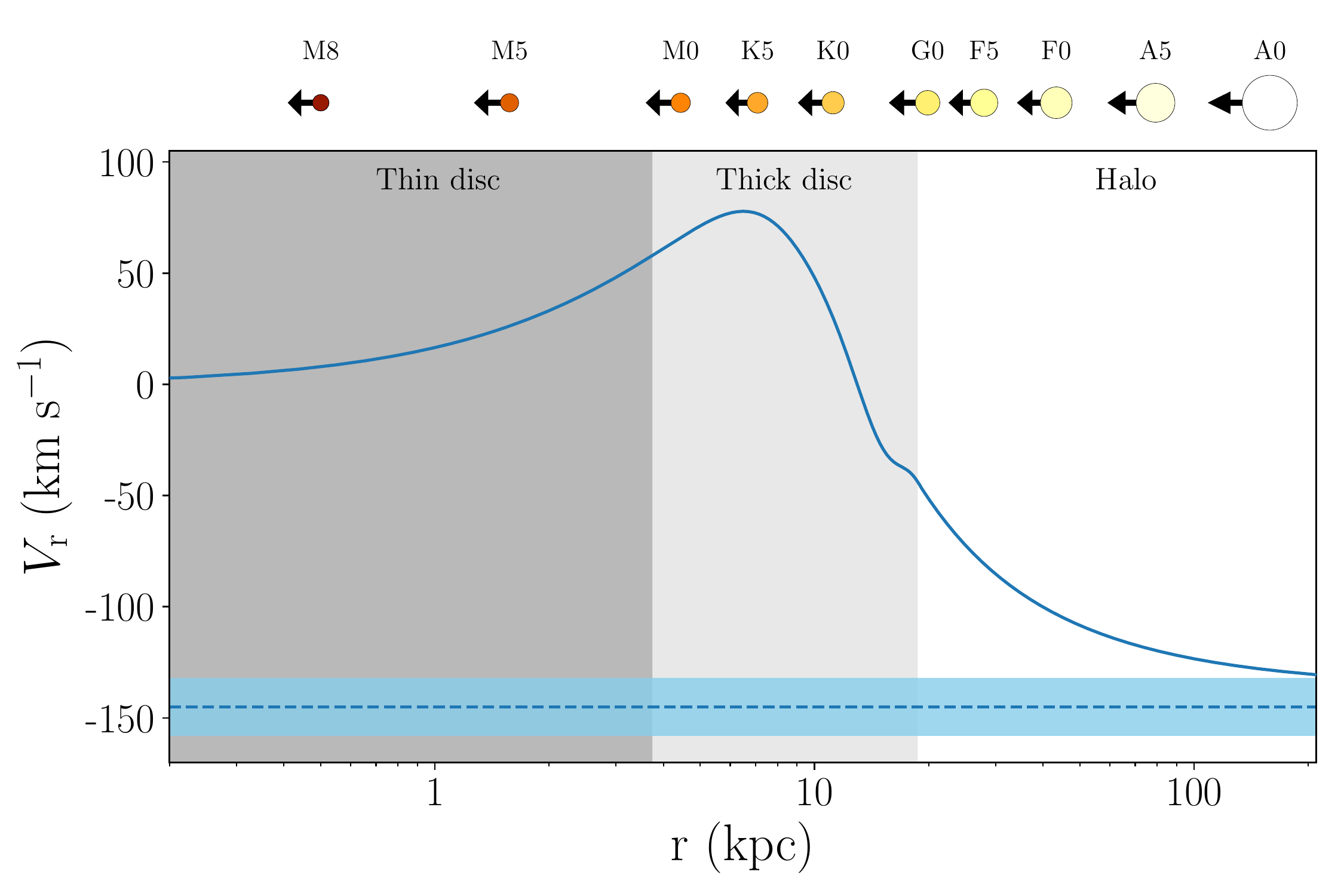}
    \caption{ Radial velocity in the direction of MAXI~J1807+132 as a function of $r$ from the rotation curve of the Galaxy of \citet[][blue solid line]{Clemens1985}. Shadowed areas indicate the extension of the thin disc ($0\leq z <1$~kpc), the thick disc ($1\leq z \leq 5$~kpc), and the halo ($h>5$ kp) projected on $r$ (Gilmore1983). The systemic velocity proposed and its corresponding error ($\gamma=-145\pm13$~km~s$^{-1}$) are indicated as a blue dashed line. Top axis shows the spectral type of a main sequence star of \textit{g}~$\sim21$ placed at a distance $r$.}
    \label{fig:discussion}
\end{figure}

We have measured a systematic velocity offset in the emission lines observed both in outburst and quiescence, that we interpret as the binary systemic velocity ($\gamma=-145\pm13$~km~s$^{-1}$). We now consider a simple kinematic model for the Milky Way to compare the observed $\gamma$ with the radial velocity expected for a given distance in the direction of the source [i.e., V$_{r}(r,l,b)$ where $r$ is the heliocentric distance]. We use the rotation curve from \citet{Clemens1985}, which is defined both in the inner and outer Galaxy, and assumes that the Milky Way follows a pure circular motion. Fig. \ref{fig:discussion}, shows the radial velocity in the direction $(l, b)=(40^{\circ}.12, 15^{\circ}.50)$ as a function of $r$. The kinetic model predicts negative radial velocities beyond $r \sim10$~kpc, while velocities consistent with the measured $\gamma$ are only expected for $r>100$~kpc. Clearly the observed systemic velocity cannot be explained by the rotation curve of the Galaxy, not even if the object is located in the very outskirts of the Galaxy. We also use $b$ to determine the vertical distance above the Galactic plane ($z$) as a function of $r$. Shadowed areas on Fig. \ref{fig:discussion} indicate the extension of the thin disc ($0\leq z <1$~kpc), the thick disc ($1\leq z \leq 5$~kpc), and the halo ($z>5$~kpc) projected on $r$ \citep{Gilmore1983}. We observe that the Galactic rotation could account for the measured radial velocity only if the source is located in the outer halo. For lower elevations the source requires a significant peculiar velocity, whether it belongs to the thin disc or to the thick disc. Such high peculiar velocities have been observed in LMXBs and are typically interpreted as natal kicks resulted from asymmetries in the supernova explosion. 

Finally, assuming that the quiescent light is dominated by the companion star, it is possible to estimate its spectral classification based on the quiescent magnitude \citep{Denisenko2017}. However, the high variability observed suggests that the disc emission is significant even in quiescence. We therefore infer an upper limit to the expected spectral type of the donor for the case of a main sequence star at a given distance (assuming \textit{g}~$\sim21$). The results are schematically presented in the top axis of the Fig. \ref{fig:discussion}

\subsection{Reflares during the decay}\label{reflares}

\begin{figure}
	\includegraphics[width=\columnwidth]{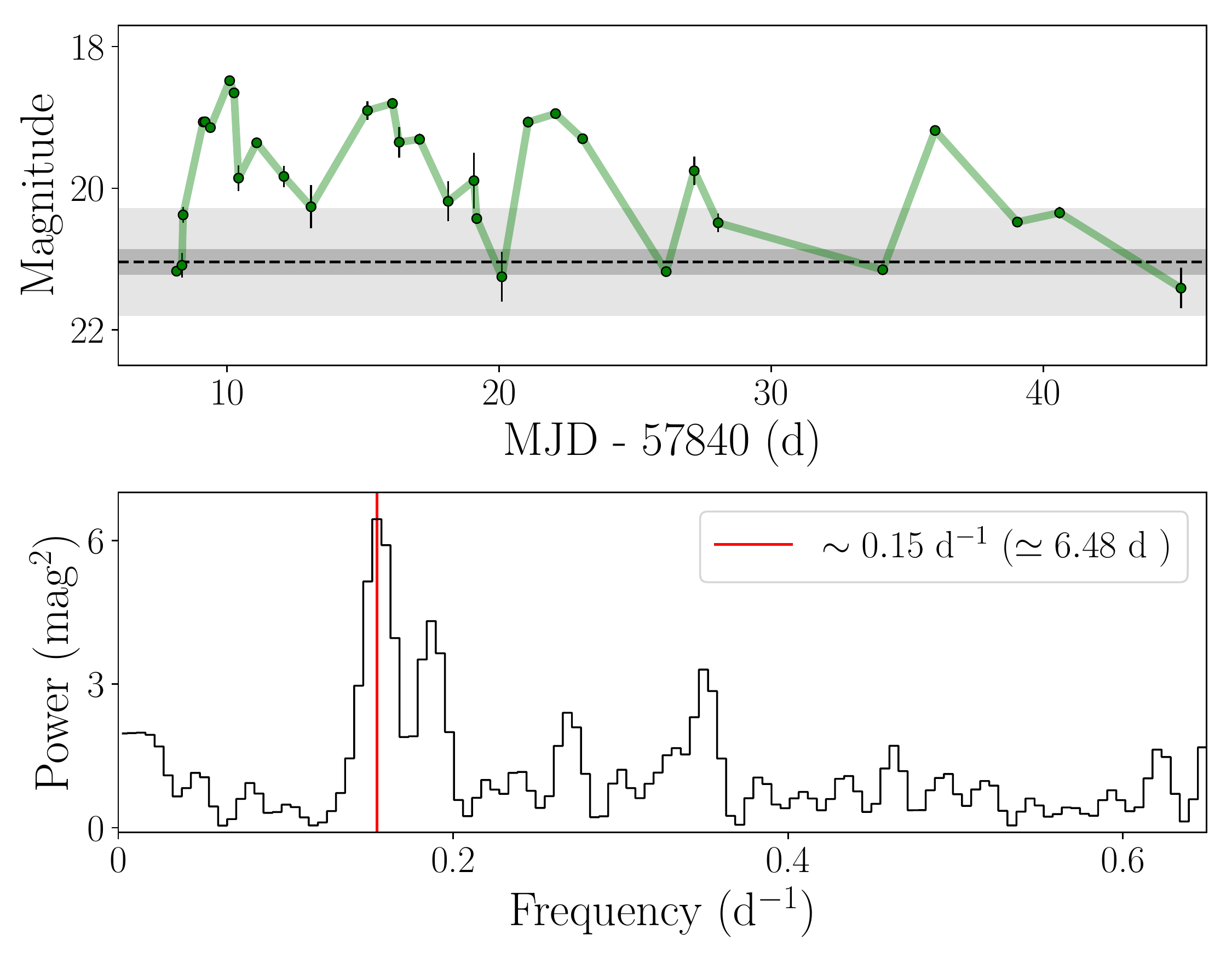}
    \caption{Top panel: Segment of the SDSS-\textit{g} light curve used for the Lomb-Scargle analysis. Bottom panel: Resulted Lomb-Scargle periodogram from the data in the top panel. The red line indicates the position of the strongest peak that corresponds to $\sim0.15$~d$^{-1}$ ($\simeq6.56$~d). }
    \label{fig:periodogram}
\end{figure}

We detected up to 7 brightening episodes during the outburst decay. A quasi-periodic recurrence can be noticed by visual inspection (Fig. \ref{fig:periodogram}). We computed a Lomb-Scargle periodogram of the SDSS-~\textit{g} points between MJD 57848--57885, when the brightening episodes are observed and resolved (i.e., sampling better than one point every 3 days). The light curve analysed and the resulting periodogram are shown in Fig. \ref{fig:periodogram}. The periodogram was produced using the \textsc{LombScargle-python} class \citep{VanderPlas2017}. We find the highest power at $\sim$ 0.15 d$^{-1}$ that yields a period of 6.5 days. The origin of these periodic brightening episodes is uncertain, but it has been proposed they might be related to a cyclical mechanism involving the accretion rate \citep{Chen1993, Augusteijn1993, Mineshige1994}. In the context of the model proposed by \citet{Augusteijn1993} the brightening episodes or reflares are "echoes" of the main outburst. X-ray illumination of the companion star triggers enhanced mass transfer and subsequent mini-outbusts. Within this picture, the recurrence time of each re-brightening is the time that a perturbation takes to travel across the accretion disc radius ($R_{\mathrm{out}}$). The heating front propagates at $v \approx \alpha v_{\mathrm{s}}$, where $ \alpha$ is the viscosity parameter and $v_{\mathrm{s}}$ the sound speed \citep{Lasota2001}. Under the assumption of a hot accretion disc \citep[$\alpha= 0.1$ and $v_{\mathrm{s}}=15$~km~s$^{-1}$,][]{Menou1999}, we obtain $R_{\mathrm{out}}\sim 8.4 \times10^{10} $ cm. Using this $R_{\mathrm{out}}$ we estimate the binary period ($P$) using $R_{\mathrm{out}}\simeq 1.2 \times 10^{11}M^{1/3}P^{2/3}$~[cm], where $M$ is the mass of the compact object expressed in $ M_{\odot}$ \citep{King1996}. We adopt the two canonical masses for compact stars, 1.4 $M_{\odot}$ for a NS and 8 $M_{\odot}$ for a BH. We obtained $P$(NS)~$\sim12$~h and $P$(BH)~$\sim5$~h, respectively. 

Using the above orbital periods, we can calculate the mean density of the companion star that fills its Roche lobe \citep{Faulkner1972}, and compare it with that of typical main sequence stars \citep{Cox2000}. We find that, for the case of a BH, the donor would be a M0--M2 main sequence while for the case of a NS it would be a F0--F5 main sequence. The distance to the system can be inferred in both cases, if we combine the lower limit to the quiescent magnitude and the spectral types above. The source is expected to be at $\gtrsim4$~kpc and $\gtrsim35$~kpc for a BH and NS system, respectively (see Fig. \ref{fig:discussion}). We note that only the BH case is consistent with the constraints devised by the optical/X-ray luminosity diagram. All in all, these broad observational properties seem to support the case of a BH binary instead of a NS. 

For simplicity, we have considered in the above a main sequence companion star. However, we note that a somewhat evolved donor cannot be ruled out. Indeed, subgiant companions are thought to be present in some LMXBs with orbital periods $\gtrapprox2$~d \citep[e.g., V404~Cygni, GX 339-4 and XTE~J1550-564;][respectively]{King1993,Munoz-Darias2008}. For instance, in the case of GX~339-4, the `stripped-giant' model predicts magnitudes in the range of $r\sim21$--24  for a distance within $\sim6$--15~kpc. This magnitude range is compatible to that observed in MAXI~1807+132 during quiescence \citep[PanSTARRS-1 broadband filters \textit{r} $=21.19\pm0.09$][]{Denisenko2017}. We also explored the possibility of a giant (luminosity class III) companion star. We assumed the observed quiescent magnitude to be the magnitude of the donor, and compared it with that expected for a field giant at given distance. Using the absolute magnitudes tabulated in \citet{Cox2000} we obtained a distance in the range of $\sim100$--180~kpc, which effectively discards a giant companion.

\subsection{On the nature of the compact object}
\label{co}
The definitive determination of the nature of the compact object in an LMXB requires of either the detection of unequivocal NS features (i.e., pulsations or thermonuclear burst) or, alternatively, a dynamical measurement of the mass of the compact object. When none of the above is available, the likely nature of the accretor can be studied based on other observables:

\begin{itemize}

\item X-ray spectral properties. Albeit both NS and BH transients share a common phenomenology during outburst (e.g., \citealt{Munoz-Darias2011a}; \citealt{Munoz-Darias2014}), the X-ray spectral modelling usually requires of an extra component to account for the emission from the NS surface \citep[or its surroundings; e.g.,][]{Lin2007,ArmasPadilla2017b,ArmasPadilla2018}. Using this argument \citet{Shidatsu2017} favoured a NS nature based on spectral fitting of the \textit{Swift} data. Here, we have also included deeper \xmm\ observations in the X-ray spectral analysis. Although we are able to fit our data with both BH-like and NS-like spectral models, some of the spectral parameters are more similar to those typically obtained for NS systems at low accretion regimes. In particular, the low  disc normalization ($N_{\rm in}=15$) and the $\sim30\%$ thermal contribution to the total flux (0.5--10~keV) challenge the BH scenario. BH systems at low luminosities reveal cold disc components, but these tend to have much larger normalization values (a proxy for the inner disc radius) and provide in most of the cases lower contributions to the observed flux \citep[e.g.,][]{Reis2010,ArmasPadilla2014a,Shidatsu2014,Plant2015}. On the other hand, several NS LMXBs accreting at low luminosities show thermal fractions consistent with those reported in this work. These have been observed in both  persistent \citep[e.g.,][]{ArmasPadilla2013,Degenaar2017,ArmasPadilla2018} and transient systems \citep[e.g.,][]{Degenaar2013,Campana2014,Arnason2015}. In addition, we can compare the $\Gamma=2.48\pm0.03$ obtained when solely using the Comptonization component with the photon index/X-ray luminosity diagram presented in \citet{Wijnands2015}, which shows a distinctive evolution for BHs and NSs at $L_{\rm X}<10^{36}~\mathrm{erg~s}^{-1}$.  Even if the distance to MAXI~J1807+132 (and therefore the luminosity) is not constrained, we find that a NS accretor is qualitatively favoured.  All in all, our X-ray fitting agree to some extent with \citet{Shidatsu2017} and favours a NS nature. However, it is important to bear in mind that a BH-like spectral modelling provides an equally good fit to the data, with the exception of the low normalization value of the disc component. Nevertheless, we note that we are using data with limited signal-to-noise and simple (Newtonian) disc models. Thus, we conclude that only from the X-ray fitting point of view we are not able to definitely constrain the nature of the compact object.  

\item The optical/X-ray luminosity diagram. This tool was initially presented by \cite{Russell2006} and has been subsequently applied to several objects \citep[e.g.,][]{ArmasPadilla2011,HernandezSantisteban2018}. This test produces a "qualitative" classification, and strongly relies on the distance to the source. In Fig. \ref{fig:OptXrayCorr} we show that a BH nature is clearly favoured for $d \gtrsim 5$~kpc, while lower distances would also allow for a NS accretor. This latter case is significantly favoured for very low distances ($d \lesssim 1$~kpc).

\item We have estimated the systemic velocity from the optical spectra (Sec. \ref{lines}). We find that it is likely large and negative ($\gamma \sim -150$~km~s$^{-1}$). Comparing this value with the rotation curve of the Galaxy (Fig. \ref{fig:discussion}) it is clear that a natal kick would be needed to explain $\gamma$, unless the system is located unrealistically far away. In addition, distances between $\sim$ 5 and 10~kpc would require very high proper motions in excess of $\sim 200$~km~s$^{-1}$. In any case, for $d\gtrsim5$~kpc (i.e., BH scenario according to the optical/X-ray ratio) the system would be located in the thick disc, which itself favours a natal kick to explain such a large height above the Galactic plane \citep{Repetto2012}. We note that a population of transient BH-LMXBs is known to be present at these high latitudes \citep[see e.g.,][and references therein]{MataSanchez2015, Kuulkers2013}. Interestingly, several of these BH systems have relatively short orbital periods \citep[see table 2 in][]{Shahbaz2013}. On the other hand, at very close distances, systemic velocities within the range observed in NS-LMXBs would be required \citep{Ramachandran1997,MataSanchez2015}. Both, a large height above the Galactic plane and a close distance agree with the low extinction along the line-of-sight detected in X-rays (Sect. \ref{Extinction}).

\item Orbital period estimates from re-flares. The presence of several optical re-flares is one of the most relevant characteristics of the outburst decay of MAXI~J1807+132. This kind of phenomena has been observed previously in cataclysmic variables \citep[e.g.,][]{Patterson1998} but also in several BH transients such as XTE~J1118+480, XTE~J1859+226, and GRO~J0422+32 \citep[see][and references therein]{Zurita2006} and NSs systems \citep[e.g.,][]{Torres2008,Patruno2016}. These binaries have orbital periods of 4.1, 6.6, and 5.1 hr, respectively \citep{Corral-Santana2016}. For the case of GRO~J0422+32 the re-flare recurrence time can be estimated and it is about 7-8 days \citep[see e.g. fig. 2 in][]{Zurita2006}. This is remarkably similar to the one we have measured in MAXI~J1807+132 (6.5 days) and from which we have inferred an orbital period of $\sim 5$ h should the compact object be a BH (Sec. \ref{reflares}). Therefore, it seems that the proposed scenario (irradiation induced changes in mass transfer) might work for GRO~J0422+32 and for extension for MAXI~J1807+132. If this is the case, the NS scenario is clearly disfavoured, as it would imply an orbital period of $\sim 12$ hours, which would place the source at $\gtrsim35$~kpc (see above). At this large distance the optical luminosity diagram strongly suggests a BH nature.

\item X-ray detections near optical quiescence. Finally, it is worth mentioning that MAXI~J1807+132 is detected in X-rays at optical magnitudes consistent with optical quiescence (e.g., grey band in Fig. \ref{fig:allphot}). This naively suggest that (i) the system might be also close to X-ray quiescence and (ii) it might be relatively nearby as truly quiescent LMXBs can be only detected within a few kpc and using deep \textit{XMM}/\textit{Chandra} observations \citep[e.g.,][and references therein]{ArmasPadilla2014b}. By stacking several \textit{Swift} observations around day 75 (see Fig. \ref{fig:allphot}) and assuming a power-law spectrum with a photon-index of 2 with $N_{\mathrm{H}}=1\times10^{-21}$ cm$^{-2}$ we obtain an unabsorbed flux of $2.8\times10^{-13}$ ergs s$^{-1}$. The combined count-rate is slightly lower than our faintest detection, so this stacking is effectively our dimmest detection of the source. By Imposing the lowest X-ray quiescence luminosities observed in LMXBs, that is, $0.5\times10^{31}$ (BH) and $5\times10^{31}$ (NS) ergs s$^{-1}$ \citep[fig. 3 in][]{ArmasPadilla2014b}, we place lower limits to the distance of 0.5 and 1.5~kpc for a BH and a NS, respectively. These values would imply very late spectral type donors, even within the brown dwarf regime (Fig. \ref{fig:discussion}). However, it seems more likely that the source is not in X-ray quiescence between re-flares as it usually takes several months to reach this state. In addition, the "optical quiescence" detected by PanSTARRs and in this work could be far from true optical quiescence. This is supported by the apparent lack of companion star features in the faintest spectra and by the flickering observed in the WHT lightcurve (Fig. \ref{fig:allphot}). We note that even if our faintest detection (see above) corresponds to luminosities as low as $1\times10^{33}$ ergs s$^{-1}$ (the quiescent luminosity shown by several BH and NS) a distance of $> 6.5$~kpc is obtained. This, combined with the optical/X-ray luminosity diagram, disfavours the NS case. 
\end{itemize}

\section{Conclusions}
We have presented an extensive observational study of the newly discovered X-ray transient MAXI~J1807+132 during its 2017 outburst decay. All the observables are consistent with those typically observed in transient LMXBs during outburst. The system displays striking properties such as dramatic changes in the optical spectrum and several re-flares with a periodicity of 6.5 days. We have explored the possible nature of the compact object, and both a NS or a BH accretor are consistent with the observations. While the NS scenario is favoured by the X-ray fitting (see also \citealt{Shidatsu2017}) it also requires fine-tuning of other parameters and some of the suggested explanations for the observed phenomenology to be wrong (e.g., re-flares). All considered, we think that MAXI~J1807+132 might be very similar to the BH transient GRO~J0422+32. Both objects have shown the same transitions between optical absorption and emission lines, as well as re-flares with very similar recurrence times. GRO~J0422+32 has an orbital period of 5.1 hr and a M4 donor \citep{Casares1995a, Webb2000,Gelino2003}, a solution very close to that we have found for MAXI~J1807+132 if the re-flares are induced by irradiation of the companion star. Both objects have the same quiescent magnitude ($r\sim 21$) but GRO~J0422+32 shows donor spectral features in the spectrum, which suggest that its distance of 2.5~kpc could be a lower limit to that of MAXI~J1807+132. However, we stress that, in any case, a NS accretor cannot be ruled out and is even favoured by the X-ray modelling. For this to be the case a distance in the range of $\sim$ 1.5--5~kpc would be required. Future observations of the source, either during new outbursts or quiescence, should provide insights on the nature of the compact object and other properties of the source.




\bibliographystyle{mnras}
\bibliography{biblio.bib} 

\begin{thebibliography}{}
\makeatletter
\relax
\def\mn@urlcharsother{\let\do\@makeother \do\$\do\&\do\#\do\^\do\_\do\%\do\~}
\def\mn@doi{\begingroup\mn@urlcharsother \@ifnextchar [ {\mn@doi@}
  {\mn@doi@[]}}
\def\mn@doi@[#1]#2{\def\@tempa{#1}\ifx\@tempa\@empty \href
  {http://dx.doi.org/#2} {doi:#2}\else \href {http://dx.doi.org/#2} {#1}\fi
  \endgroup}
\def\mn@eprint#1#2{\mn@eprint@#1:#2::\@nil}
\def\mn@eprint@arXiv#1{\href {http://arxiv.org/abs/#1} {{\tt arXiv:#1}}}
\def\mn@eprint@dblp#1{\href {http://dblp.uni-trier.de/rec/bibtex/#1.xml}
  {dblp:#1}}
\def\mn@eprint@#1:#2:#3:#4\@nil{\def\@tempa {#1}\def\@tempb {#2}\def\@tempc
  {#3}\ifx \@tempc \@empty \let \@tempc \@tempb \let \@tempb \@tempa \fi \ifx
  \@tempb \@empty \def\@tempb {arXiv}\fi \@ifundefined
  {mn@eprint@\@tempb}{\@tempb:\@tempc}{\expandafter \expandafter \csname
  mn@eprint@\@tempb\endcsname \expandafter{\@tempc}}}

\bibitem[\protect\citeauthoryear{Armas~Padilla, Degenaar, Patruno, Russell,
  Linares, Maccarone, Homan  \& Wijnands}{Armas~Padilla
  et~al.}{2011}]{ArmasPadilla2011}
Armas~Padilla M.,  Degenaar N.,  Patruno A.,  Russell D.~M.,  Linares M.,
  Maccarone T.~J.,  Homan J.,   Wijnands R.,  2011, \mn@doi [\mnras]
  {10.1111/j.1365-2966.2011.19308.x}, 417, 659

\bibitem[\protect\citeauthoryear{Armas~Padilla, Wijnands  \&
  Degenaar}{Armas~Padilla et~al.}{2013}]{ArmasPadilla2013}
Armas~Padilla M.,  Wijnands R.,   Degenaar N.,  2013, \mn@doi [\mnras]
  {10.1093/mnrasl/slt119}, 436, L89

\bibitem[\protect\citeauthoryear{Armas~Padilla, Wijnands, Altamirano,
  M{\'e}ndez, Miller  \& Degenaar}{Armas~Padilla
  et~al.}{2014a}]{ArmasPadilla2014a}
Armas~Padilla M.,  Wijnands R.,  Altamirano D.,  M{\'e}ndez M.,  Miller J.~M.,
   Degenaar N.,  2014a, \mn@doi [\mnras] {10.1093/mnras/stu243}, 439, 3908

\bibitem[\protect\citeauthoryear{Armas~Padilla, Wijnands, Degenaar,
  Mu{\~n}oz-Darias, Casares  \& Fender}{Armas~Padilla
  et~al.}{2014b}]{ArmasPadilla2014b}
Armas~Padilla M.,  Wijnands R.,  Degenaar N.,  Mu{\~n}oz-Darias T.,  Casares
  J.,   Fender R.~P.,  2014b, \mn@doi [\mnras] {10.1093/mnras/stu1487}, 444,
  902

\bibitem[\protect\citeauthoryear{Armas~Padilla, Ueda, Hori, Shidatsu  \&
  Mu{\~n}oz-Darias}{Armas~Padilla et~al.}{2017a}]{ArmasPadilla2017b}
Armas~Padilla M.,  Ueda Y.,  Hori T.,  Shidatsu M.,   Mu{\~n}oz-Darias T.,
  2017a, \mn@doi [\mnras] {10.1093/mnras/stx020}, 467, 290

\bibitem[\protect\citeauthoryear{Armas~Padilla, Wijnands, Degenaar,
  Munoz-Darias, Jimenez-Ibarra, Mata~Sanchez, Casares  \&
  Charles}{Armas~Padilla et~al.}{2017b}]{ArmasPadilla2017a}
Armas~Padilla M.,  Wijnands R.,  Degenaar N.,  Munoz-Darias T.,  Jimenez-Ibarra
  F.,  Mata~Sanchez D.,  Casares J.,   Charles P.~A.,  2017b, The Astronomer's
  Telegram, 10224

\bibitem[\protect\citeauthoryear{Armas~Padilla, Ponti, De~Marco,
  Mu{\~n}oz-Darias  \& Haberl}{Armas~Padilla et~al.}{2018}]{ArmasPadilla2018}
Armas~Padilla M.,  Ponti G.,  De~Marco B.,  Mu{\~n}oz-Darias T.,   Haberl F.,
  2018, \mn@doi [\mnras] {10.1093/mnras/stx2538}, 473, 3789

\bibitem[\protect\citeauthoryear{Arnason, Sivakoff, Heinke, Cohn  \&
  Lugger}{Arnason et~al.}{2015}]{Arnason2015}
Arnason R.~M.,  Sivakoff G.~R.,  Heinke C.~O.,  Cohn H.~N.,   Lugger P.~M.,
  2015, \mn@doi [\apj] {10.1088/0004-637X/807/1/52}, 807, 52

\bibitem[\protect\citeauthoryear{Arnaud}{Arnaud}{1996}]{Arnaud1996}
Arnaud K.~A.,  1996, in {Jacoby} G.~H.,  {Barnes} J.,  eds,  Astronomical
  Society of the Pacific Conference Series Vol. 101, Astronomical Data Analysis
  Software and Systems V. p.~17, \url
  {http://adsabs.harvard.edu/abs/1996ASPC..101...17A}

\bibitem[\protect\citeauthoryear{{Astropy Collaboration} et~al.,}{{Astropy
  Collaboration} et~al.}{2013}]{Astropy2013}
{Astropy Collaboration} et~al., 2013, \mn@doi [\aap]
  {10.1051/0004-6361/201322068}, \href
  {http://adsabs.harvard.edu/abs/2013A%26A...558A..33A} {558, A33}

\bibitem[\protect\citeauthoryear{Augusteijn, Kuulkers  \& Shaham}{Augusteijn
  et~al.}{1993}]{Augusteijn1993}
Augusteijn T.,  Kuulkers E.,   Shaham J.,  1993, \aap, 279, L13

\bibitem[\protect\citeauthoryear{Bailyn \& Orosz}{Bailyn \&
  Orosz}{1995}]{Bailyn1995}
Bailyn C.~D.,  Orosz J.~A.,  1995, \mn@doi [\apjl] {10.1086/187764}, 440, L73

\bibitem[\protect\citeauthoryear{Belloni, Motta  \& Mu{\~n}oz-Darias}{Belloni
  et~al.}{2011}]{Belloni2011}
Belloni T.~M.,  Motta S.~E.,   Mu{\~n}oz-Darias T.,  2011, Bulletin of the
  Astronomical Society of India, 39, 409

\bibitem[\protect\citeauthoryear{Burke, Gilfanov  \& Sunyaev}{Burke
  et~al.}{2017}]{Burke2017}
Burke M.~J.,  Gilfanov M.,   Sunyaev R.,  2017, \mn@doi [\mnras]
  {10.1093/mnras/stw2514}, 466, 194

\bibitem[\protect\citeauthoryear{Burrows et~al.,}{Burrows
  et~al.}{2005}]{Burrows2005}
Burrows D.~N.,  et~al., 2005, \mn@doi [\ssr] {10.1007/s11214-005-5097-2}, 120,
  165

\bibitem[\protect\citeauthoryear{Campana, Brivio, Degenaar, Mereghetti,
  Wijnands, D'Avanzo, Israel  \& Stella}{Campana et~al.}{2014}]{Campana2014}
Campana S.,  Brivio F.,  Degenaar N.,  Mereghetti S.,  Wijnands R.,  D'Avanzo
  P.,  Israel G.~L.,   Stella L.,  2014, \mn@doi [\mnras]
  {10.1093/mnras/stu709}, 441, 1984

\bibitem[\protect\citeauthoryear{Casares \& Jonker}{Casares \&
  Jonker}{2014}]{Casares2014}
Casares J.,  Jonker P.~G.,  2014, \mn@doi [\ssr] {10.1007/s11214-013-0030-6},
  183, 223

\bibitem[\protect\citeauthoryear{Casares, Charles  \& Naylor}{Casares
  et~al.}{1992}]{Casares1992}
Casares J.,  Charles P.~A.,   Naylor T.,  1992, \mn@doi [\nat]
  {10.1038/355614a0}, 355, 614

\bibitem[\protect\citeauthoryear{Casares, Marsh, Charles, Martin, Martin,
  Harlaftis, Pavlenko  \& Wagner}{Casares et~al.}{1995}]{Casares1995a}
Casares J.,  Marsh T.~R.,  Charles P.~A.,  Martin A.~C.,  Martin E.~L.,
  Harlaftis E.~T.,  Pavlenko E.~P.,   Wagner R.~M.,  1995, \mn@doi [\mnras]
  {10.1093/mnras/274.2.565}, 274, 565

\bibitem[\protect\citeauthoryear{Cepa et~al.,}{Cepa et~al.}{2000}]{Cepa2000}
Cepa J.,  et~al., 2000, in {Iye} M.,  {Moorwood} A.~F.,  eds,  \procspie Vol.
  4008, Optical and IR Telescope Instrumentation and Detectors. pp 623--631,
  \mn@doi{10.1117/12.395520}, \url
  {http://adsabs.harvard.edu/abs/2000SPIE.4008..623C}

\bibitem[\protect\citeauthoryear{Chambers et~al.,}{Chambers
  et~al.}{2016}]{Chambers2016}
Chambers K.~C.,  et~al., 2016, preprint (\mn@eprint {arXiv} {1612.05560})

\bibitem[\protect\citeauthoryear{Charles \& Coe}{Charles \&
  Coe}{2006}]{Charles2006}
Charles P.~A.,  Coe M.~J.,  2006, Optical, ultraviolet and infrared
  observations of X-ray binaries.
pp 215--265, \url {http://adsabs.harvard.edu/abs/2006csxs.book..215C}

\bibitem[\protect\citeauthoryear{Chen, Livio  \& Gehrels}{Chen
  et~al.}{1993}]{Chen1993}
Chen W.,  Livio M.,   Gehrels N.,  1993, \mn@doi [\apjl] {10.1086/186817}, 408,
  L5

\bibitem[\protect\citeauthoryear{Clemens}{Clemens}{1985}]{Clemens1985}
Clemens D.~P.,  1985, \mn@doi [\apj] {10.1086/163386}, 295, 422

\bibitem[\protect\citeauthoryear{Corral-Santana, Casares, Mu{\~{n}}oz-Darias,
  Bauer, Mart{\'{\i}}nez-Pais  \& Russell}{Corral-Santana
  et~al.}{2016}]{Corral-Santana2016}
Corral-Santana J.~M.,  Casares J.,  Mu{\~{n}}oz-Darias T.,  Bauer F.~E.,
  Mart{\'{\i}}nez-Pais I.~G.,   Russell D.~M.,  2016, \mn@doi [Astronomy {\&}
  Astrophysics] {10.1051/0004-6361/201527130}, 587, A61

\bibitem[\protect\citeauthoryear{Cox}{Cox}{2000}]{Cox2000}
Cox A.~N.,  2000, Allen's astrophysical quantities.
\url {http://adsabs.harvard.edu/abs/2000asqu.book.....C}

\bibitem[\protect\citeauthoryear{Degenaar, Wijnands  \& Miller}{Degenaar
  et~al.}{2013}]{Degenaar2013}
Degenaar N.,  Wijnands R.,   Miller J.~M.,  2013, \mn@doi [\apjl]
  {10.1088/2041-8205/767/2/L31}, 767, L31

\bibitem[\protect\citeauthoryear{Degenaar, Pinto, Miller, Wijnands, Altamirano,
  Paerels, Fabian  \& Chakrabarty}{Degenaar et~al.}{2017}]{Degenaar2017}
Degenaar N.,  Pinto C.,  Miller J.~M.,  Wijnands R.,  Altamirano D.,  Paerels
  F.,  Fabian A.~C.,   Chakrabarty D.,  2017, \mn@doi [\mnras]
  {10.1093/mnras/stw2355}, 464, 398

\bibitem[\protect\citeauthoryear{Denisenko}{Denisenko}{2017}]{Denisenko2017}
Denisenko D.,  2017, The Astronomer's Telegram, 10217

\bibitem[\protect\citeauthoryear{D{\'{\i}}az~Trigo, Parmar, Boirin, M{\'e}ndez
  \& Kaastra}{D{\'{\i}}az~Trigo et~al.}{2006}]{DiazTrigo2006}
D{\'{\i}}az~Trigo M.,  Parmar A.~N.,  Boirin L.,  M{\'e}ndez M.,   Kaastra
  J.~S.,  2006, \mn@doi [\aap] {10.1051/0004-6361:20053586}, 445, 179

\bibitem[\protect\citeauthoryear{Dubus, Kim, Menou, Szkody  \& Bowen}{Dubus
  et~al.}{2001}]{Dubus2001}
Dubus G.,  Kim R. S.~J.,  Menou K.,  Szkody P.,   Bowen D.~V.,  2001, \mn@doi
  [\apj] {10.1086/320648}, 553, 307

\bibitem[\protect\citeauthoryear{Faulkner, Flannery  \& Warner}{Faulkner
  et~al.}{1972}]{Faulkner1972}
Faulkner J.,  Flannery B.~P.,   Warner B.,  1972, \mn@doi [\apjl]
  {10.1086/180989}, 175, L79

\bibitem[\protect\citeauthoryear{Fender \& Mu{\~{n}}oz-Darias}{Fender \&
  Mu{\~{n}}oz-Darias}{2016}]{Fender2016}
Fender R.,  Mu{\~{n}}oz-Darias T.,  2016, in , Astrophysical Black Holes.
Springer International Publishing, pp 65--100,
  \mn@doi{10.1007/978-3-319-19416-5_3}

\bibitem[\protect\citeauthoryear{Foight, G{\"u}ver, {\"O}zel  \& Slane}{Foight
  et~al.}{2016}]{Foight2016}
Foight D.~R.,  G{\"u}ver T.,  {\"O}zel F.,   Slane P.~O.,  2016, \mn@doi [\apj]
  {10.3847/0004-637X/826/1/66}, 826, 66

\bibitem[\protect\citeauthoryear{Galloway, Muno, Hartman, Psaltis  \&
  Chakrabarty}{Galloway et~al.}{2008}]{Galloway2008}
Galloway D.~K.,  Muno M.~P.,  Hartman J.~M.,  Psaltis D.,   Chakrabarty D.,
  2008, \mn@doi [\apjs] {10.1086/592044}, 179, 360

\bibitem[\protect\citeauthoryear{Gehrels et~al.,}{Gehrels
  et~al.}{2004}]{Gehrels2004}
Gehrels N.,  et~al., 2004, \mn@doi [\apj] {10.1086/422091}, 611, 1005

\bibitem[\protect\citeauthoryear{Gelino \& Harrison}{Gelino \&
  Harrison}{2003}]{Gelino2003}
Gelino D.~M.,  Harrison T.~E.,  2003, \mn@doi [\apj] {10.1086/379311}, 599,
  1254

\bibitem[\protect\citeauthoryear{Gilmore \& Reid}{Gilmore \&
  Reid}{1983}]{Gilmore1983}
Gilmore G.,  Reid N.,  1983, \mn@doi [\mnras] {10.1093/mnras/202.4.1025}, 202,
  1025

\bibitem[\protect\citeauthoryear{Hern{\'a}ndez~Santisteban, Knigge, Pretorius,
  Sullivan  \& Warner}{Hern{\'a}ndez~Santisteban
  et~al.}{2018}]{HernandezSantisteban2018}
Hern{\'a}ndez~Santisteban J.~V.,  Knigge C.,  Pretorius M.~L.,  Sullivan M.,
  Warner B.,  2018, \mn@doi [\mnras] {10.1093/mnras/stx2296}, 473, 3241

\bibitem[\protect\citeauthoryear{Jansen et~al.,}{Jansen
  et~al.}{2001}]{Jansen2001}
Jansen F.,  et~al., 2001, \mn@doi [\aap] {10.1051/0004-6361:20000036}, 365, L1

\bibitem[\protect\citeauthoryear{Jim{\'{e}}nez-Ibarra, Mu{\~{n}}oz-Darias,
  Wang, Casares, S{\'{a}}nchez, Steeghs, Padilla  \&
  Charles}{Jim{\'{e}}nez-Ibarra et~al.}{2018}]{Jimenez-Ibarra2018}
Jim{\'{e}}nez-Ibarra F.,  Mu{\~{n}}oz-Darias T.,  Wang L.,  Casares J.,
  S{\'{a}}nchez D.~M.,  Steeghs D.,  Padilla M.~A.,   Charles P.~A.,  2018,
  \mn@doi [Monthly Notices of the Royal Astronomical Society]
  {10.1093/mnras/stx2926}, 474, 4717

\bibitem[\protect\citeauthoryear{Kalberla, Burton, Hartmann, Arnal, Bajaja,
  Morras  \& Pöppel}{Kalberla et~al.}{2005}]{Kalberla2005}
Kalberla P. M.~W.,  Burton W.~B.,  Hartmann D.,  Arnal E.~M.,  Bajaja E.,
  Morras R.,   Pöppel W. G.~L.,  2005, \mn@doi [Astronomy {\&} Astrophysics]
  {10.1051/0004-6361:20041864}, 440, 775

\bibitem[\protect\citeauthoryear{Kennea, Evans, Beardmore, Krimm, Romano,
  Yamaoka, Serino  \& Negoro}{Kennea et~al.}{2017a}]{Kennea2017a}
Kennea J.~A.,  Evans P.~A.,  Beardmore A.~P.,  Krimm H.~A.,  Romano P.,
  Yamaoka K.,  Serino M.,   Negoro H.,  2017a, The Astronomer's Telegram, 10215

\bibitem[\protect\citeauthoryear{Kennea et~al.,}{Kennea
  et~al.}{2017b}]{Kennea2017b}
Kennea J.~A.,  et~al., 2017b, The Astronomer's Telegram, 10216

\bibitem[\protect\citeauthoryear{Kennea et~al.,}{Kennea
  et~al.}{2017c}]{Kennea2017}
Kennea J.~A.,  et~al., 2017c, The Astronomer's Telegram, 10216

\bibitem[\protect\citeauthoryear{Khargharia, Froning, Robinson  \&
  Gelino}{Khargharia et~al.}{2012}]{Khargharia2012}
Khargharia J.,  Froning C.~S.,  Robinson E.~L.,   Gelino D.~M.,  2012, \mn@doi
  [The Astronomical Journal] {10.1088/0004-6256/145/1/21}, 145, 21

\bibitem[\protect\citeauthoryear{King}{King}{1993}]{King1993}
King A.~R.,  1993, \mn@doi [\mnras] {10.1093/mnras/260.1.L5}, 260, L5

\bibitem[\protect\citeauthoryear{King, Kolb  \& Burderi}{King
  et~al.}{1996}]{King1996}
King A.~R.,  Kolb U.,   Burderi L.,  1996, \mn@doi [\apjl] {10.1086/310105},
  464, L127

\bibitem[\protect\citeauthoryear{Kubota, Tanaka, Makishima, Ueda, Dotani, Inoue
   \& Yamaoka}{Kubota et~al.}{1998}]{Kubota1998}
Kubota A.,  Tanaka Y.,  Makishima K.,  Ueda Y.,  Dotani T.,  Inoue H.,
  Yamaoka K.,  1998, \mn@doi [\pasj] {10.1093/pasj/50.6.667}, 50, 667

\bibitem[\protect\citeauthoryear{Kuulkers et~al.,}{Kuulkers
  et~al.}{2013}]{Kuulkers2013}
Kuulkers E.,  et~al., 2013, \mn@doi [\aap] {10.1051/0004-6361/201219447}, 552,
  A32

\bibitem[\protect\citeauthoryear{Lasota}{Lasota}{2001}]{Lasota2001}
Lasota J.-P.,  2001, \mn@doi [New Astronomy Reviews]
  {10.1016/s1387-6473(01)00112-9}, 45, 449

\bibitem[\protect\citeauthoryear{Lin, Remillard  \& Homan}{Lin
  et~al.}{2007}]{Lin2007}
Lin D.,  Remillard R.~A.,   Homan J.,  2007, \mn@doi [The Astrophysical
  Journal] {10.1086/521181}, 667, 1073

\bibitem[\protect\citeauthoryear{Makishima, Maejima, Mitsuda, Bradt, Remillard,
  Tuohy, Hoshi  \& Nakagawa}{Makishima et~al.}{1986}]{Makishima1986}
Makishima K.,  Maejima Y.,  Mitsuda K.,  Bradt H.~V.,  Remillard R.~A.,  Tuohy
  I.~R.,  Hoshi R.,   Nakagawa M.,  1986, \mn@doi [\apj] {10.1086/164534}, 308,
  635

\bibitem[\protect\citeauthoryear{Mata~S{\'a}nchez, Mu{\~n}oz-Darias, Casares,
  Corral-Santana  \& Shahbaz}{Mata~S{\'a}nchez et~al.}{2015}]{MataSanchez2015}
Mata~S{\'a}nchez D.,  Mu{\~n}oz-Darias T.,  Casares J.,  Corral-Santana J.~M.,
   Shahbaz T.,  2015, \mn@doi [\mnras] {10.1093/mnras/stv2111}, 454, 2199

\bibitem[\protect\citeauthoryear{Menou, Hameury  \& Stehle}{Menou
  et~al.}{1999}]{Menou1999}
Menou K.,  Hameury J.-M.,   Stehle R.,  1999, \mn@doi [Monthly Notices of the
  Royal Astronomical Society] {10.1046/j.1365-8711.1999.02396.x}, 305, 79

\bibitem[\protect\citeauthoryear{Miller, Raymond, Fabian, Steeghs, Homan,
  Reynolds, van~der Klis  \& Wijnands}{Miller et~al.}{2006}]{Miller2006}
Miller J.~M.,  Raymond J.,  Fabian A.,  Steeghs D.,  Homan J.,  Reynolds C.,
  van~der Klis M.,   Wijnands R.,  2006, \mn@doi [\nat] {10.1038/nature04912},
  441, 953

\bibitem[\protect\citeauthoryear{Mineshige}{Mineshige}{1994}]{Mineshige1994}
Mineshige S.,  1994, \mn@doi [\apjl] {10.1086/187482}, 431, L99

\bibitem[\protect\citeauthoryear{Mitsuda et~al.,}{Mitsuda
  et~al.}{1984}]{Mitsuda1984}
Mitsuda K.,  et~al., 1984, \pasj, 36, 741

\bibitem[\protect\citeauthoryear{Motta}{Motta}{2016}]{Motta2016}
Motta S.~E.,  2016, \mn@doi [Astronomische Nachrichten]
  {10.1002/asna.201612320}, 337, 398

\bibitem[\protect\citeauthoryear{Munari \& Zwitter}{Munari \&
  Zwitter}{1997}]{Munari1997}
Munari U.,  Zwitter T.,  1997, \mn@doi [\aap] {10.1051/aas:1998386}, 318, 269

\bibitem[\protect\citeauthoryear{Mu{\~n}oz-Darias, Casares  \&
  Mart{\'{\i}}nez-Pais}{Mu{\~n}oz-Darias et~al.}{2008}]{Munoz-Darias2008}
Mu{\~n}oz-Darias T.,  Casares J.,   Mart{\'{\i}}nez-Pais I.~G.,  2008, \mn@doi
  [\mnras] {10.1111/j.1365-2966.2008.12987.x}, 385, 2205

\bibitem[\protect\citeauthoryear{Mu{\~n}oz-Darias, Motta  \&
  Belloni}{Mu{\~n}oz-Darias et~al.}{2011}]{Munoz-Darias2011a}
Mu{\~n}oz-Darias T.,  Motta S.,   Belloni T.~M.,  2011, \mn@doi [\mnras]
  {10.1111/j.1365-2966.2010.17476.x}, 410, 679

\bibitem[\protect\citeauthoryear{Mu{\~n}oz-Darias, Fender, Motta  \&
  Belloni}{Mu{\~n}oz-Darias et~al.}{2014}]{Munoz-Darias2014}
Mu{\~n}oz-Darias T.,  Fender R.~P.,  Motta S.~E.,   Belloni T.~M.,  2014,
  \mn@doi [\mnras] {10.1093/mnras/stu1334}, 443, 3270

\bibitem[\protect\citeauthoryear{Mu{\~n}oz-Darias et~al.,}{Mu{\~n}oz-Darias
  et~al.}{2016}]{Munoz-Darias2016}
Mu{\~n}oz-Darias T.,  et~al., 2016, \mn@doi [\nat] {10.1038/nature17446}, 534,
  75

\bibitem[\protect\citeauthoryear{Munoz-Darias, Jimenez-Ibarra, Mata~Sanchez,
  Armas~Padilla, Casares  \& Charles}{Munoz-Darias
  et~al.}{2017}]{Munoz-Darias2017a}
Munoz-Darias T.,  Jimenez-Ibarra F.,  Mata~Sanchez D.,  Armas~Padilla M.,
  Casares J.,   Charles P.~A.,  2017, The Astronomer's Telegram, 10221

\bibitem[\protect\citeauthoryear{Negoro et~al.,}{Negoro
  et~al.}{2017}]{Negoro2017}
Negoro H.,  et~al., 2017, The Astronomer's Telegram, 10208

\bibitem[\protect\citeauthoryear{Patruno, Maitra, Curran, D'Angelo,
  Fridriksson, Russell, Middleton  \& Wijnands}{Patruno
  et~al.}{2016}]{Patruno2016}
Patruno A.,  Maitra D.,  Curran P.~A.,  D'Angelo C.,  Fridriksson J.~K.,
  Russell D.~M.,  Middleton M.,   Wijnands R.,  2016, \mn@doi [\apj]
  {10.3847/0004-637X/817/2/100}, 817, 100

\bibitem[\protect\citeauthoryear{Patterson et~al.,}{Patterson
  et~al.}{1998}]{Patterson1998}
Patterson J.,  et~al., 1998, \mn@doi [\pasp] {10.1086/316252}, 110, 1290

\bibitem[\protect\citeauthoryear{Plant, Fender, Ponti, Mu{\~n}oz-Darias  \&
  Coriat}{Plant et~al.}{2015}]{Plant2015}
Plant D.~S.,  Fender R.~P.,  Ponti G.,  Mu{\~n}oz-Darias T.,   Coriat M.,
  2015, \mn@doi [\aap] {10.1051/0004-6361/201423925}, 573, A120

\bibitem[\protect\citeauthoryear{Ponti, Fender, Begelman, Dunn, Neilsen  \&
  Coriat}{Ponti et~al.}{2012}]{Ponti2012}
Ponti G.,  Fender R.~P.,  Begelman M.~C.,  Dunn R. J.~H.,  Neilsen J.,   Coriat
  M.,  2012, \mn@doi [\mnras] {10.1111/j.1745-3933.2012.01224.x}, 422, L11

\bibitem[\protect\citeauthoryear{Ramachandran \& Bhattacharya}{Ramachandran \&
  Bhattacharya}{1997}]{Ramachandran1997}
Ramachandran R.,  Bhattacharya D.,  1997, \mn@doi [\mnras]
  {10.1093/mnras/288.3.565}, 288, 565

\bibitem[\protect\citeauthoryear{Reis, Fabian  \& Miller}{Reis
  et~al.}{2010}]{Reis2010}
Reis R.~C.,  Fabian A.~C.,   Miller J.~M.,  2010, \mn@doi [\mnras]
  {10.1111/j.1365-2966.2009.15976.x}, 402, 836

\bibitem[\protect\citeauthoryear{Remillard \& McClintock}{Remillard \&
  McClintock}{2006}]{Remillard2006}
Remillard R.~A.,  McClintock J.~E.,  2006, \mn@doi [Annual Review of Astronomy
  and Astrophysics] {10.1146/annurev.astro.44.051905.092532}, 44, 49

\bibitem[\protect\citeauthoryear{Repetto, Davies  \& Sigurdsson}{Repetto
  et~al.}{2012}]{Repetto2012}
Repetto S.,  Davies M.~B.,   Sigurdsson S.,  2012, \mn@doi [\mnras]
  {10.1111/j.1365-2966.2012.21549.x}, 425, 2799

\bibitem[\protect\citeauthoryear{Russell, Fender, Hynes, Brocksopp, Homan,
  Jonker  \& Buxton}{Russell et~al.}{2006}]{Russell2006}
Russell D.~M.,  Fender R.~P.,  Hynes R.~I.,  Brocksopp C.,  Homan J.,  Jonker
  P.~G.,   Buxton M.~M.,  2006, \mn@doi [\mnras]
  {10.1111/j.1365-2966.2006.10756.x}, 371, 1334

\bibitem[\protect\citeauthoryear{Russell, Fender  \& Jonker}{Russell
  et~al.}{2007}]{Russell2007}
Russell D.~M.,  Fender R.~P.,   Jonker P.~G.,  2007, \mn@doi [\mnras]
  {10.1111/j.1365-2966.2007.12008.x}, 379, 1108

\bibitem[\protect\citeauthoryear{Shahbaz, Russell, Zurita, Casares,
  Corral-Santana, Dhillon  \& Marsh}{Shahbaz et~al.}{2013}]{Shahbaz2013}
Shahbaz T.,  Russell D.~M.,  Zurita C.,  Casares J.,  Corral-Santana J.~M.,
  Dhillon V.~S.,   Marsh T.~R.,  2013, \mn@doi [\mnras]
  {10.1093/mnras/stt1212}, 434, 2696

\bibitem[\protect\citeauthoryear{Shidatsu et~al.,}{Shidatsu
  et~al.}{2014}]{Shidatsu2014}
Shidatsu M.,  et~al., 2014, \mn@doi [\apj] {10.1088/0004-637X/789/2/100}, 789,
  100

\bibitem[\protect\citeauthoryear{Shidatsu et~al.,}{Shidatsu
  et~al.}{2017a}]{Shidatsu2017}
Shidatsu M.,  et~al., 2017a, preprint (\mn@eprint {arXiv} {1710.03371})

\bibitem[\protect\citeauthoryear{Shidatsu et~al.,}{Shidatsu
  et~al.}{2017b}]{Shidatsu2017a}
Shidatsu M.,  et~al., 2017b, The Astronomer's Telegram, 10222

\bibitem[\protect\citeauthoryear{Str{\"u}der et~al.,}{Str{\"u}der
  et~al.}{2001}]{Strueder2001}
Str{\"u}der L.,  et~al., 2001, \mn@doi [\aap] {10.1051/0004-6361:20000066},
  365, L18

\bibitem[\protect\citeauthoryear{Tomsick, Yamaoka, Corbel, Kaaret, Kalemci  \&
  Migliari}{Tomsick et~al.}{2009}]{Tomsick2009}
Tomsick J.~A.,  Yamaoka K.,  Corbel S.,  Kaaret P.,  Kalemci E.,   Migliari S.,
   2009, \mn@doi [\apjl] {10.1088/0004-637X/707/1/L87}, 707, L87

\bibitem[\protect\citeauthoryear{Torres et~al.,}{Torres
  et~al.}{2008}]{Torres2008}
Torres M. A.~P.,  et~al., 2008, \mn@doi [\apj] {10.1086/523831}, 672, 1079

\bibitem[\protect\citeauthoryear{Turner et~al.,}{Turner
  et~al.}{2001}]{Turner2001}
Turner M. J.~L.,  et~al., 2001, \mn@doi [\aap] {10.1051/0004-6361:20000087},
  365, L27

\bibitem[\protect\citeauthoryear{VanderPlas}{VanderPlas}{2017}]{VanderPlas2017}
VanderPlas J.~T.,  2017, preprint (\mn@eprint {arXiv} {1703.09824})

\bibitem[\protect\citeauthoryear{Verner, Ferland, Korista  \& Yakovlev}{Verner
  et~al.}{1996}]{Verner1996}
Verner D.~A.,  Ferland G.~J.,  Korista K.~T.,   Yakovlev D.~G.,  1996, \mn@doi
  [\apj] {10.1086/177435}, 465, 487

\bibitem[\protect\citeauthoryear{Webb, Naylor, Ioannou, Charles  \&
  Shahbaz}{Webb et~al.}{2000}]{Webb2000}
Webb N.~A.,  Naylor T.,  Ioannou Z.,  Charles P.~A.,   Shahbaz T.,  2000,
  \mn@doi [\mnras] {10.1046/j.1365-8711.2000.03608.x}, 317, 528

\bibitem[\protect\citeauthoryear{Wijnands, Degenaar, Armas~Padilla, Altamirano,
  Cavecchi, Linares, Bahramian  \& Heinke}{Wijnands
  et~al.}{2015}]{Wijnands2015}
Wijnands R.,  Degenaar N.,  Armas~Padilla M.,  Altamirano D.,  Cavecchi Y.,
  Linares M.,  Bahramian A.,   Heinke C.~O.,  2015, \mn@doi [\mnras]
  {10.1093/mnras/stv1974}, 454, 1371

\bibitem[\protect\citeauthoryear{Wilms, Allen  \& McCray}{Wilms
  et~al.}{2000}]{Wilms2000}
Wilms J.,  Allen A.,   McCray R.,  2000, \mn@doi [\apj] {10.1086/317016}, 542,
  914

\bibitem[\protect\citeauthoryear{Zdziarski, Johnson  \& Magdziarz}{Zdziarski
  et~al.}{1996}]{Zdziarski1996}
Zdziarski A.~A.,  Johnson W.~N.,   Magdziarz P.,  1996, \mn@doi [\mnras]
  {10.1093/mnras/283.1.193}, 283, 193

\bibitem[\protect\citeauthoryear{Zurita et~al.,}{Zurita
  et~al.}{2002}]{Zurita2002}
Zurita C.,  et~al., 2002, \mn@doi [Monthly Notices of the Royal Astronomical
  Society] {10.1046/j.1365-8711.2002.05588.x}, 334, 999

\bibitem[\protect\citeauthoryear{Zurita et~al.,}{Zurita
  et~al.}{2006}]{Zurita2006}
Zurita C.,  et~al., 2006, \mn@doi [\apj] {10.1086/503286}, 644, 432

\bibitem[\protect\citeauthoryear{{\.Z}ycki, Done  \& Smith}{{\.Z}ycki
  et~al.}{1999}]{Zycki1999}
{\.Z}ycki P.~T.,  Done C.,   Smith D.~A.,  1999, \mn@doi [\mnras]
  {10.1046/j.1365-8711.1999.02885.x}, 309, 561

\bibitem[\protect\citeauthoryear{van Dokkum}{van Dokkum}{2001}]{Dokkum2001}
van Dokkum P.~G.,  2001, \mn@doi [\pasp] {10.1086/323894}, 113, 1420

\bibitem[\protect\citeauthoryear{van~der Klis}{van~der
  Klis}{2006}]{vanderKlis2006}
van~der Klis M.,  2006, Rapid X-ray Variability.
pp 39--112, \url {http://adsabs.harvard.edu/abs/2006csxs.book...39V}

\makeatother
\end{thebibliography}



\appendix

\section{Optical photometry of MAXI~J1807+132 in the 2017 outburst}

The photometric data presented in this work are listed in Table \ref{tab:journald_phot}. They correspond to the SDSS-\textit{g},-\textit{r}, and -\textit{i} bands. The telescope used in each case is indicated.

\begin{table*}
	\centering
	\caption{SDSS-\textit{g}, -\textit{r}, and -\textit{i} photometry of MAXI~J1807+132 (LT, LCO, and GTC). }
	\label{tab:journald_phot}
	\begin{tabular}{lccccccc} 
		\hline
		 &   \multicolumn{3}{c}{SDSS-\textit{g} (mag)}&  \multicolumn{2}{c}{SDSS-\textit{r} (mag)}& \multicolumn{2}{c}{SDSS-\textit{i} (mag)} \\
		 Date & LT & LCO&GTC  &LT&LCO&LT&LCO\\
		\hline
28-03-2017 & - & 18.76 $\pm$ 0.10 & 18.36 $\pm$ 0.01 & - & 18.54 $\pm$ 0.07 & - & 18.59 $\pm$ 0.10 \\
30-03-2017 & 19.24 $\pm$ 0.02 & - & 19.22 $\pm$ 0.02 & 19.24 $\pm$ 0.02 & - & 19.26 $\pm$ 0.02 & - \\
31-03-2017 & 19.75 $\pm$ 0.03 & - & - & 19.70 $\pm$ 0.04 & - & 19.79 $\pm$ 0.04 & - \\
01-04-2017 & 20.17 $\pm$ 0.04 & - & - & 20.10 $\pm$ 0.04 & - & 20.05 $\pm$ 0.06 & - \\
02-04-2017 & 21.14 $\pm$ 0.04 & - & - & 20.98 $\pm$ 0.04 & - & 20.77 $\pm$ 0.04 & - \\
03-04-2017 & 20.63 $\pm$ 0.09 & - & - & 20.78 $\pm$ 0.11 & - & 20.88 $\pm$ 0.17 & - \\
04-04-2017 & 20.89 $\pm$ 0.06 & - & - & 20.62 $\pm$ 0.04 & - & 20.61 $\pm$ 0.06 & - \\
05-04-2017 & 21.17 $\pm$ 0.03 & 21.09 $\pm$ 0.17 & - & 21.08 $\pm$ 0.03 & 20.69 $\pm$ 0.14 & 20.90 $\pm$ 0.04 & 20.67 $\pm$ 0.17 \\
06-04-2017 & 19.06 $\pm$ 0.02 & 19.14 $\pm$ 0.04 & 19.06 $\pm$ 0.02 & 19.09 $\pm$ 0.02 & 19.22 $\pm$ 0.04 & 19.15 $\pm$ 0.01 & 19.41 $\pm$ 0.06 \\
07-04-2017 & 18.48 $\pm$ 0.04 & 19.86 $\pm$ 0.18 & 18.65 $\pm$ 0.05 & 18.54 $\pm$ 0.03 & 19.74 $\pm$ 0.15 & 18.38 $\pm$ 0.02 & 20.18 $\pm$ 0.26 \\
08-04-2017 & 19.36 $\pm$ 0.06 & - & - & 19.38 $\pm$ 0.03 & - & 19.47 $\pm$ 0.03 & - \\
09-04-2017 & 19.83 $\pm$ 0.15 & - & - & 19.81 $\pm$ 0.10 & - & 19.89 $\pm$ 0.08 & - \\
10-04-2017 & 20.26 $\pm$ 0.31 & - & - & 20.44 $\pm$ 0.32 & - & 20.28 $\pm$ 0.24 & - \\
12-04-2017 & 18.90 $\pm$ 0.13 & - & - & 19.17 $\pm$ 0.10 & - & 19.27 $\pm$ 0.08 & - \\
13-04-2017 & 18.80 $\pm$ 0.07 & 19.35 $\pm$ 0.21 & - & - & 19.66 $\pm$ 0.26 & 18.82 $\pm$ 0.04 & 19.27 $\pm$ 0.24 \\
14-04-2017 & 19.31 $\pm$ 0.08 & - & - & - & 19.52 $\pm$ 0.29 & 19.42 $\pm$ 0.07 & 19.37 $\pm$ 0.28 \\
15-04-2017 & 20.18 $\pm$ 0.28 & - & - & - & - & 19.74 $\pm$ 0.12 & - \\
16-04-2017 & 20.43 $\pm$ 0.07 & 19.89 $\pm$ 0.39 & - & - & 20.41 $\pm$ 0.69 & 20.09 $\pm$ 0.08 & 19.21 $\pm$ 0.37 \\
17-04-2017 & 21.25 $\pm$ 0.35 & - & - & - & - & - & - \\
18-04-2017 & 19.07 $\pm$ 0.07 & - & - & - & - & 18.92 $\pm$ 0.08 & - \\
19-04-2017 & 18.95 $\pm$ 0.02 & - & - & - & - & 18.80 $\pm$ 0.02 & - \\
20-04-2017 & 19.30 $\pm$ 0.02 & - & - & - & - & - & - \\
23-04-2017 & 21.18 $\pm$ 0.03 & - & - & - & - & - & - \\
24-04-2017 & - & 19.75 $\pm$ 0.20 & - & - & 20.70 $\pm$ 0.40 & - & 21.04 $\pm$ 0.75 \\
25-04-2017 & - & 20.49 $\pm$ 0.13 & - & - & 20.10 $\pm$ 0.13 & - & 20.33 $\pm$ 0.17 \\
01-05-2017 & 21.15 $\pm$ 0.03 & - & - & - & - & - & - \\
03-05-2017 & 19.18 $\pm$ 0.04 & - & - & - & - & - & - \\
06-05-2017 & 20.48 $\pm$ 0.05 & - & - & - & - & - & - \\
07-05-2017 & - & 20.35 $\pm$ 0.08 & - & - & 20.66 $\pm$ 0.08 & - & 20.32 $\pm$ 0.10 \\
12-05-2017 & 21.41 $\pm$ 0.29 & - & - & - & - & - & 21.05 $\pm$ 0.65 \\
16-05-2017 & - & - & - & - & 19.10 $\pm$ 0.65 & - & - \\
31-05-2017 & 21.03 $\pm$ 0.03 & - & - & - & - & - & - \\
04-06-2017 & - & 19.49 $\pm$ 0.19 & - & - & 19.46 $\pm$ 0.11 & - & 20.82 $\pm$ 0.44 \\
19-06-2017 & 21.27 $\pm$ 0.03 & - & - & - & - & - & - \\
12-07-2017 & 21.25 $\pm$ 0.03 & - & - & - & - & - & - \\
16-07-2017 & - & - & 21.54 $\pm$ 0.08 & - & - & - & - \\
18-08-2017 & - & - & 21.47 $\pm$ 0.04 & - & - & - & - \\

		\hline
	\end{tabular}
\end{table*}

\section*{Acknowledgements}

We are grateful to the anonymous referee for useful comments and suggestions that have improved the paper. We acknowledge support by the Spanish MINECO under grant AYA2017-83216-P. Based on observations made with the Gran Telescopio Canarias (GTC), instaled in the Spanish Observatorio del Roque de los Muchachos of the Instituto de Astrof\'{i}sica de Canarias, in the island of La Palma. The authors are thankful to the GTC team that carried out the ToO observations. We are thankful to Phil Charles, Nathalie Degenaar, Rudy Wijnands, and Tom Maccarone for useful discussion on the nature of the system. The WHT is operated on the island of La Palma by the Isaac Newton Group of Telescopes in the Spanish Observatorio del Roque de los Muchachos of the Instituto de Astrof\'{i}sica de Canarias. Based on observations made with the Liverpool Telescope operated on the island of La Palma by Liverpool John Moores University in the Spanish Observatorio del Roque de los Muchachos of the Instituto de Astrof\'{i}sica de Canarias with financial support from the UK Science and Technology Facilities Council. TMD is supported by RYC-2015-18148. MAP's research is funded under the Juan de la Cierva Fellowship Programme (IJCI-2016-30867) from MINECO. MAPT acknowledges support via a Ram\'on y Cajal Fellowship (RYC-2015-17854).  D.M.S. acknowledges support from the ERC under the European Union's Horizon 2020 research and innovation programme (grant agreement No. 715051; Spiders). \textsc{molly} software developed by T. R. Marsh is gratefully acknowledged. The Faulkes Telescope Project is an education partner of LCO. The Faulkes Telescopes are maintained and operated by LCO.

\bsp	
\label{lastpage}
\end{document}